\documentclass[preprint,longbibliography,
superscriptaddress,
 amsmath,amssymb,
 aps,floatfix
]{revtex4-1}
\usepackage{gensymb}
\usepackage{xcolor}
\usepackage{graphicx}% Include figure files
\usepackage{dcolumn}% Align table columns on the decimal point
\usepackage{bm}% bold math
\usepackage{comment}
\usepackage{multirow}
\usepackage{textcomp, gensymb}
\usepackage{soul}
\usepackage{
}
\newcommand{\Cr}{CsCr$_3$Sb$_5$}
\newcommand{\V}{V$_3$Sb$_5$}
\newcommand{\g}{$\Gamma$}

\newcommand{\tc}{$T_{\rm CDW}$}
\newcommand{\ef}{$E_F$}

\newcommand{\kz}{$k_z$}

\DeclareUnicodeCharacter{2212}{-}
\begin{document}

\title{Spin Excitations and Flat Electronic Bands in a Cr-based Kagome Superconductor}

\author{Zehao Wang*}
\affiliation{Department of Physics and Astronomy, Rice University, Houston, 77005 Texas, USA}

\author{Yucheng Guo*}
\affiliation{Department of Physics and Astronomy, Rice University, Houston, 77005 Texas, USA}

\author{Hsiao-Yu Huang*}
\affiliation{National Synchrotron Radiation Research Center, Hsinchu 300092, Taiwan}

\author{Fang Xie*}
\affiliation{Department of Physics and Astronomy, Rice University, Houston, 77005 Texas, USA}

\author{Yuefei Huang}
\affiliation{Department of Materials Science and NanoEngineering, Rice University, Houston, Texas 77005, USA}

\author{Bin Gao}
\affiliation{Department of Physics and Astronomy, Rice University, Houston, 77005 Texas, USA}%

\author{Ji Seop Oh}
\affiliation{Department of Physics and Astronomy, Rice University, Houston, 77005 Texas, USA}%
\affiliation{Department of Physics, University of California, Berkeley, 94720 California, USA}%
\affiliation{Department of Applied Physics, Sookmyung Women’s University, Seoul 04310, Republic of Korea}
\affiliation{Institute of Advanced Materials and Systems, Sookmyung Women’s University, Seoul 04310, Republic of Korea}

\author{Han Wu}
\affiliation{Department of Physics and Astronomy, Rice University, Houston, 77005 Texas, USA}%

\author{Jun Okamoto}
\affiliation{National Synchrotron Radiation Research Center, Hsinchu 300092, Taiwan}

\author{Ganesha Channagowdra}
\affiliation{National Synchrotron Radiation Research Center, Hsinchu 300092, Taiwan}

\author{Chien-Te Chen}
\affiliation{National Synchrotron Radiation Research Center, Hsinchu 300092, Taiwan}

\author{Feng Ye}
\affiliation{Neutron Scattering Division, Oak Ridge National Laboratory, Oak Ridge, TN 37831, USA}

\author{Xingye Lu}
\affiliation{Center for Advanced Quantum Studies, School of Physics and Astronomy, Beijing Normal University, Beijing 100875, China}

\author{Zhaoyu Liu}
\affiliation{Department of Physics, University of Washington,  98105 Washington, USA}%

\author{Zheng Ren}
\affiliation{Department of Physics and Astronomy, Rice University, Houston, 77005 Texas, USA}%

\author{Yuan Fang}
\affiliation{Department of Physics and Astronomy, Rice University, Houston, 77005 Texas, USA}%

\author{Yiming Wang}
\affiliation{Department of Physics and Astronomy, Rice University, Houston, 77005 Texas, USA}%

\author{Ananya Biswas}
\affiliation{Department of Physics and Astronomy, Rice University, Houston, 77005 Texas, USA}%

\author{Yichen Zhang}
\affiliation{Department of Physics and Astronomy, Rice University, Houston, 77005 Texas, USA}%

\author{Ziqin Yue}
\affiliation{Department of Physics and Astronomy, Rice University, Houston, 77005 Texas, USA}%
\affiliation{Applied Physics Graduate Program, Smalley-Curl Institute, Rice University, Houston, 77005 Texas, USA}%

\author{Cheng Hu}
\affiliation{Advanced Light Source, Lawrence Berkeley National Laboratory, Berkeley, 94720 California, USA}

\author{Chris Jozwiak}
\affiliation{Advanced Light Source, Lawrence Berkeley National Laboratory, Berkeley, 94720 California, USA}

\author{Aaron Bostwick}
\affiliation{Advanced Light Source, Lawrence Berkeley National Laboratory, Berkeley, 94720 California, USA}%

\author{Eli Rotenberg}
\affiliation{Advanced Light Source, Lawrence Berkeley National Laboratory, Berkeley, 94720 California, USA}%

\author{Makoto Hashimoto}
\affiliation{Stanford Synchrotron Radiation Lightsource, SLAC National Accelerator Laboratory, Menlo Park, 94025 California, USA}%

\author{Donghui Lu}
\affiliation{Stanford Synchrotron Radiation Lightsource, SLAC National Accelerator Laboratory, Menlo Park, 94025 California, USA}%

\author{Junichiro Kono}
\affiliation{Department of Physics and Astronomy, Rice University, Houston, 77005 Texas, USA}
\affiliation{Department of Electrical and Computer Engineering, Rice University, Houston, 77005 Texas, USA}
\affiliation{Smalley-Curl Institute, Rice University, Houston, 77005 Texas, USA}%
\affiliation{Department of Materials Science and NanoEngineering, Rice University, Houston, 77005 Texas, USA}%

\author{Jiun-Haw Chu}
\affiliation{Department of Physics, University of Washington,  98105 Washington, USA}%

\author{Boris I. Yakobson}
\affiliation{Department of Materials Science and NanoEngineering, Rice University, Houston, Texas 77005, USA}

\author{Robert J. Birgeneau}
\affiliation{Department of Physics, University of California, Berkeley, 94720 California, USA}

\author{Guang-Han Cao}
\affiliation{School of Physics, Zhejiang University, Hangzhou 310058, China}

\author{Atsushi Fujimori}
\affiliation{National Synchrotron Radiation Research Center, Hsinchu 300092, Taiwan}
\affiliation{Center for Quantum Science and Technology and Department of Physics,
National Tsing Hua University, Hsinchu 300044, Taiwan
}
\affiliation{Department of Physics, University of Tokyo, Bunkyo-Ku, Tokyo 113-0033, Japan}

\author{Di-Jing Huang}
\email{djhuang@nsrrc.org.tw}
\affiliation{National Synchrotron Radiation Research Center, Hsinchu 300092, Taiwan}
\affiliation{Department of Physics, National Tsing Hua University, Hsinchu 300044, Taiwan}
\affiliation{Department of Electrophysics, National Yang Ming Chiao Tung University, Hsinchu 300093, Taiwan}

\author{Qimiao Si}
\email{qmsi@rice.edu}
\affiliation{Department of Physics and Astronomy, Rice University, Houston, 77005 Texas, USA}%
\affiliation{Smalley-Curl Institute, Rice University, Houston, 77005 Texas, USA}%

\author{Ming Yi}
\email{mingyi@rice.edu}
\affiliation{Department of Physics and Astronomy, Rice University, Houston, 77005 Texas, USA}%
\affiliation{Smalley-Curl Institute, Rice University, Houston, 77005 Texas, USA}%

\author{Pengcheng Dai}
\email{pdai@rice.edu}
\affiliation{Department of Physics and Astronomy, Rice University, Houston, 77005 Texas, USA}%
\affiliation{Smalley-Curl Institute, Rice University, Houston, 77005 Texas, USA}%

\date{\today}% 

\begin{abstract}
In the quest for topology- and correlation-driven quantum states, kagome lattice materials have garnered significant interest for their band structures, featuring flat bands (FBs) from the quantum destructive interference of the electronic wavefunction. Tuning an FB to the chemical potential could induce electronic instabilities and emergent orders. Despite extensive studies, direct evidence of FBs tuned to the chemical potential and their role in emergent orders in bulk materials remains lacking. Using angle-resolved photoemission spectroscopy, resonant inelastic X-ray scattering, and density functional theory, we show that the low-energy structure of the Cr-based kagome metal superconductor {\Cr} is dominated by FBs at the Fermi level. We also observe low-energy magnetic excitations evolving across the low-temperature transition, largely consistent with the FB shift. Our results suggest that the low-temperature order contains a magnetic origin and that the kagome FBs may play a role in the emergence of this order.
\end{abstract}

\maketitle

\section{Introduction}
Quantum materials with a large density of states such as associated with flat bands (FBs) can display exotic quantum states. The FBs, when tuned to near the chemical potential, can provide a large amount of degenerate electronic states across the Brillouin zone (BZ) that are available to respond to interactions, hence leading to electronic instabilities and potential emergent electronic orders, such as magnetic order, charge-density-wave, and unconventional superconductivity \cite{Checkelsky2024-zc, Das2021-ef, Xie2021-fc, Lisi2020-jf, Kirchner2020-xc, Fernandes2022-tq,Iglovikov2014-jd}. While FBs near the Fermi level can be achieved through moire superlattices of magic-angle twisted bilayer graphene \cite{Bistritzer2011-qh,Cao2018,Tian2023-ed}, geometrically frustrated lattices such as the kagome and pyrochlore systems can also exhibit FBs from quantum interference of the electronic wavefunctions \cite{Sutherland1986-vu, Lieb1989, Essafi2017, Regnault2022, Calugaru2022}. However, stabilizing a kagome lattice to bring these FBs into close proximity to the \ef~has been challenging, and establishing a relationship between FBs and the electronic/magnetic order has been particularly difficult.
% Huang_arXiv_2023, Wakefield2023-kf, Huang2024-aw

For example, extensive experimental studies on the kagome lattice materials, including the binary 11 systems ((Fe,Co)Sn~\cite{Ye2018-iy, Kang2020-gy, Liu2020-po, Kang2020-wq, Lin2020-ns} and FeGe~\cite{Teng2022-ru, Yin2022-el, Teng2023-ii}), the 135 systems (A{\V}~\cite{Ortiz2019-hh, Jiang2021-we, Li2021-kp, Ortiz2021-zp, Zhao2021-ii, Hu2022-xa, Kato2022-ow, Zheng2022-qw} and {ATi$_{3}$Bi$_{5}$} (A=K, Rb, Cs)~\cite{Werhahn2022,Yang_arXiv_2022,Liu2023-pr,Hu2023-za}), and the 166 systems ({RMn$_{6}$Sn$_{6}$} and {RV$_{6}$Sn$_{6}$} (R=rare earth)~\cite{VENTURINI199135, Yin2020-ai, Li2021-ll, Peng2021-ig, Rosenberg2022-do,Cheng2023-hp,Lee2024-bc, Hu2024-my,Cheng2024-vn}), have revealed emergent orders, from quantum magnetism and unconventional superconductivity to nematicity and charge orders. However, these phases are often discussed in connection to characteristic features in the electronic structure such as the Van Hove Singularities (VHSs) or Dirac fermions, attributed to the inherent topology of the kagome lattice,
and not to the kagome FBs~\cite{Wu2007-bs, Mielke1991-hs, Tasaki1992-ls, Miyahara2007-dw, Zhang2010-qs, Aoki2020-qm, Tang2011-jb, Bergman2008-xx}. A noteworthy development in this context is the discovery of Ni$_3$In, which is reported to host a partial FB at the {\ef} by density functional theory (DFT) predictions and simultaneously exhibit non-Fermi liquid transport behaviors~\cite{Ye2024-in}. This finding bolsters the hypothesis that a FB at the {\ef} extending throughout the momentum space could be a promising avenue to realize exotic quantum orders in bulk quantum materials. Theoretically, models of $d$-electron systems on kagome and related lattices have been mapped to a Kondo lattice description through the notion of compact molecular orbitals, through which a phase diagram with emergent flat bands and strange metallicity has been identified~\cite{HuSciAdv2023,Chen_emergent_2024,Chen_arXiv_2023}.

The chromium-based kagome metal {\Cr} has recently been discovered and identified as a promising candidate~\cite{Liu2023-vk}. Notably, this material exhibits phase transitions characterized by the appearance of superlattice peaks observed by X-ray diffraction (XRD) below \tc~$= 54$ K, suggested to be a unidirectional charge density wave (CDW) order. Magnetic susceptibility and nuclear magnetic resonance measurements indicate the simultaneous onset of a magnetic order~\cite{Liu2023-vk}. First principle calculations reveal multiple competing density wave phases \cite{Xu2023-pn, Wu2024}. Under the application of hydrostatic pressure, the two orders are separated in temperature and suppressed, and a superconducting dome appears with a $T_c$~peaking at 6.4 K \cite{Liu2023-vk}. Although the phase diagram is reminiscent of that of other known unconventional superconducting families such as the iron-based superconductors and cuprates where superconductivity competes with static magnetic order~\cite{Fernandes2022-tq, Armitage2010-px}, there has not yet been understanding of the nature of the competing order nor evidence of FBs and their association with magnetism across the 54 K phase transition.

By combining polarization-dependent angle-resolved photoemission spectroscopy (ARPES) measurements and DFT calculations, we unveil the presence of FBs near \ef~extending through a large portion of the BZ. As the temperature is lowered across \tc, this FB exhibits a downward shift of approximately 20 meV away from the {\ef}. From resonant inelastic X-ray scattering (RIXS) experiments, we find clear non-dispersive magnetic excitations above the density wave transition temperature near the BZ center. The overall energy spread as well as the center energy position of the magnetic excitations show an increase below \tc, consistent with the observed shift in FBs from the ARPES measurements. Therefore, our combined ARPES and RIXS measurements signal that the low temperature order contains a component that is magnetic in origin and that the associated excitations include electron-hole excitations of FBs, implying that FBs in the vicinity of the Fermi level play a role in the magnetic excitations in proximity to unconventional superconductivity~\cite{Fernandes2022-tq,RevModPhys.87.855,RevModPhys.84.1383,Si23.1}.

%High-quality single crystals of {\Cr}, synthesized via the flux method, 
\section{Results}
\begin{figure*}
\includegraphics[width=\textwidth]{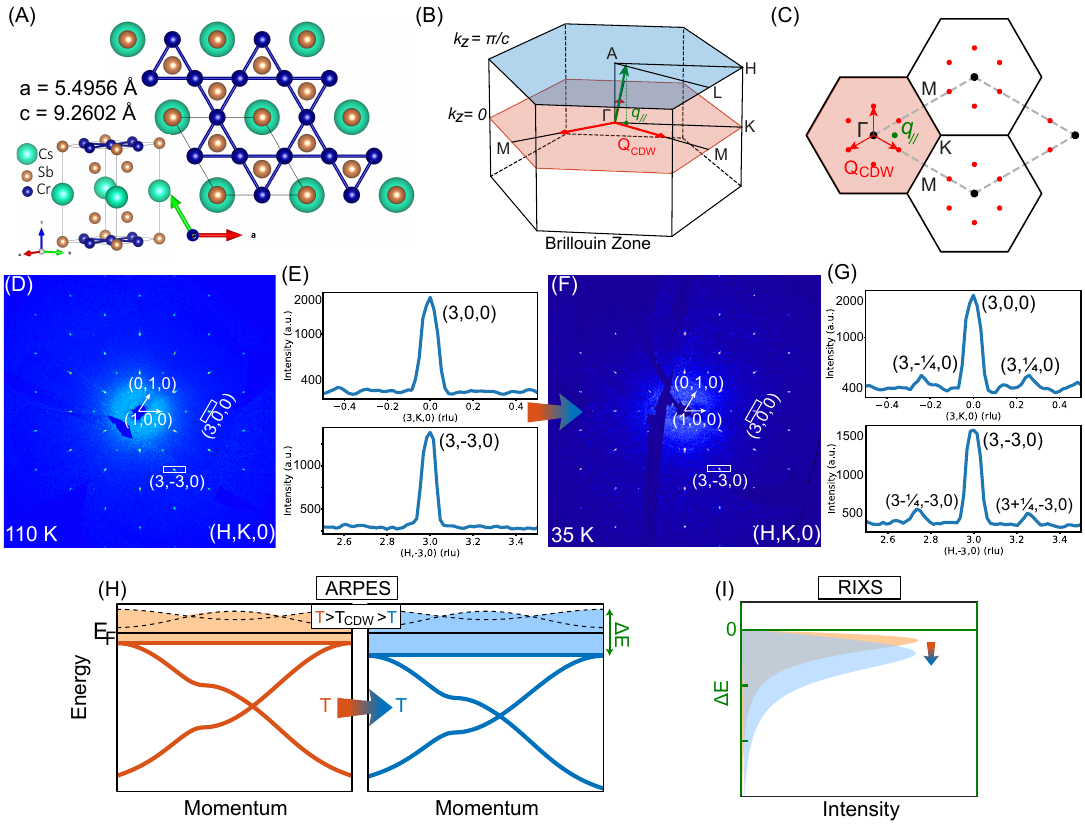}
\caption{\label{fig:fig1} \textbf{Crystal structure, X-ray diffraction, Schematic of ARPES and RIXS results.} (\textbf{A}) Unit cell of \Cr. The Cr forms a kagome lattice. (\textbf{B}) and (\textbf{C}) Reciprocal space of \Cr. CDW observed by XRD are shown as red arrows or red points. The green arrow marks the $q$ of temperature-dependent RIXS, where $q_\parallel$ is the projection of $q$ onto the sample surface. (\textbf{D}) and (\textbf{F}) XRD in the {$(H,K,0)$} plane at 110 K and 35 K. (\textbf{E}) and (\textbf{G}) Corresponding cuts of (F) and (G), respectively. (\textbf{H}) Cartoon illustration of the ARPES observed shift of the kagome flat band away from \ef~below \tc. (\textbf{I}) Cartoon illustration of the observed shift of the magnetic excitations observed by RIXS across \tc. The light orange and light blue shaded areas indicate the spectral weight 
%corresponding 
that is coupled to particle-hole excitations across \ef~illustrated in (H).  }
\end{figure*}

\Cr~crystallizes in a layered hexagonal lattice consisting of alternatingly stacked Cr-Sb sheets and Cs layers (space group P6/mmm No. 191) with lattice parameters \(a = 5.4956(1) \, \rm \AA\) and \(c = 9.2602(2) \, \rm \AA\) at 293 K, where the Cr atoms form a kagome lattice (Fig.~\ref{fig:fig1}A)~\cite{Liu2023-vk}. The characterization of \Cr~can be found in Fig. S1. The corresponding 3D BZ is plotted in Fig.~\ref{fig:fig1}B.  Figure ~\ref{fig:fig1}C shows a map of the reciprocal space in the $(H,K,0)$ plane where superlattice peaks associated with the density wave order below \tc~are marked. Our X-ray diffraction maps at 110 K (Fig.~\ref{fig:fig1}D) and 35 K (Fig.~\ref{fig:fig1}F) confirm the appearance of the superlattice peaks at $(1/4,0,0)$ and $(0,1/4,0)$ in the $(H,K,0)$ plane below \tc~\cite{Liu2023-vk}, as can be directly observed along the line cuts (Figs.~\ref{fig:fig1}E,G). We summarize our key findings in Figs.~\ref{fig:fig1}H-I. From ARPES measurements, we find FBs just $\sim$60 meV below the Fermi level that exhibit a $\sim$20 meV downward shift below \tc~(Fig.~\ref{fig:fig1}H). Our RIXS measurements conclusively identify magnetic excitations near the BZ center with a spectral width of $\sim$100 meV. These excitations broaden and shift in energy below \tc~(Fig.~\ref{fig:fig1}I), corresponding to a broadening of the magnetic electron-hole excitations due to the temperature-dependent shift of the FBs away from the Fermi level. In the following, we present our detailed ARPES and RIXS experimental results.

As {\Cr} is isostructural to the well-studied Cs{\V}, we first compare the DFT calculated band structure of the two systems. As Cr has one additional valence electron, \Cr~has more electron filling than Cs\V. This is reflected in the calculated band structure (Fig.~\ref{fig:fig2}A-B), where the main features in the two systems are qualitatively similar except an overall raised chemical potential in \Cr~compared to that of Cs\V. In particular, the kagome FBs in {\Cr} are much closer to \ef, positioned approximately 200 meV above the {\ef}, in contrast to around 1 eV for Cs{\V}. Also, the VHSs lie well below \(E_F\) for {\Cr}, while Cs{\V} hosts multiple VHSs near \ef. The difference in the predicted FB position is also pronounced in the density of states (DOS) calculations, highlighted by red arrows in Fig.~\ref{fig:fig2}C. To visualize the electronic structure of {\Cr}, we present ARPES results measured at 10 K on a kagome termination dominant surface (see core level measurement in the SM and Fig. S2). 
%From a detailed photon-energy dependence measurement, we can determine the inner potential of \Cr~to be 19 eV (see Supplementary Information and Fig. S4). 
The Fermi surface map corresponding to the k$_z$ = 0 plane is shown in comparison to that calculated from DFT (Fig.~\ref{fig:fig2}D). Noticeably, the Fermi surfaces contain a large pocket around the {\g} point and small pockets around the M points, in good alignment with the DFT calculations for the phase without the density wave order. To note, no remarkable band folding was observed on the Fermi surface. Meanwhile, the VHSs at the M points are observed at 0.5 eV below {\ef}, evident in the characteristic triangular pockets around the K points with their corners touching at the M points, giving a good overall agreement with the DFT calculations (Fig.~\ref{fig:fig2}E).

To further visualize the electronic structure of {\Cr} and understand its orbital textures, we present a detailed comparison between band dispersions along high symmetry directions and the orbital-projected DFT in Fig.~\ref{fig:fig2}. The low energy electronic states are mostly populated by the five Cr-3$d$ orbitals and Sb-5$p$ orbitals illustrated in Fig.~\ref{fig:fig2}F. We adopt the site-dependent local coordinates as the basis shown by the colored arrows to fully respect the lattice symmetry (see SM Fig. S3). Orbital-projected DFT calculations of the band structure along the high symmetry directions suggest that the topological flat bands associated with the kagome lattice are predominantly of ${d_{xz}}$ and ${d_{yz}}$ orbital character, immediately above \ef~(see SM Fig. S3). Experimentally, we can utilize the polarization dependent photoemission matrix elements to gain information on the dominant orbital character of the observed dispersions. Based on a detailed analysis of such matrix elements for all of our measurement geometries (see SM Fig. S3), we can overlay the orbital-projected calculated bands from symmetry-allowed orbitals along two high symmetry directions in the BZ under two polarization directions (Fig.~\ref{fig:fig2}), while raw data without DFT is provided in the SM Fig. S6). While most features exhibit reasonable overall agreements, we notice that there is a portion of a flat dispersion between $\bar{K}$-$\bar{M}$ near $-1$ eV (dotted white arrow in Fig.~\ref{fig:fig2}I) connected to an electron band towards $\bar{\Gamma}$. The orbital selective rules suggest these features can be attributed to ${d_{yz}}$ or ${d_{xy}}$ orbital characters (see SM Fig. S3). However, no bands in the DFT calculations match this flat portion. To capture this feature, we note that if a renormalization factor of 1.4 is applied to the ${d_{yz}}$ character band marked by the solid white arrow, the agreement can be better established (see SM Fig. S4). Another possibility that we cannot rule out is that the flat portion of ${d_{xy}}$ character near $-2$ eV being strongly renormalized. In either scenario, orbital-dependent correlation effects would need to be invoked in \Cr, which is not uncommonly found in multi-orbital systems including the iron-based superconductors~\cite{Yi2013-bz, Huang2022-ih, Yi2015-qj, Huang2023-he}, a recently reported Mn-based kagome system~\cite{Samanta2024}, as well as Ca$_{1-x}$Sr$_{x}$RuO$_4$~\cite{Neupane2009}. 

Having discussed the overall electronic structure, we next focus on the FB near \ef. First, from the orbital-projected band structure calculated by DFT, we see that the kagome topological flat band appears at 0.25 eV above {\ef}. In Fig.~\ref{fig:fig2}K-M we present the measured band dispersions within 1 eV of \ef~along \g-M measured under different polarizations together with the energy distribution curves (EDC). Interestingly for both polarizations, a flat band feature close to {\ef} can be observed. This can be seen in the ubiquitous peak in the EDCs within 100 meV of \ef, and is especially clear at the $\bar{\Gamma}$ point marked by the red arrows, where according to the DFT calculations, should only have an electron band near \ef~(Fig.~\ref{fig:fig2}A). The location of the electron band bottom can be seen in the EDCs at $\bar{\Gamma}$ in the form of a hump near -0.7 eV. The only feature above the electron band in the DFT calculation is the kagome flat band, situated slightly above \ef. Hence, if we assume that the DFT results are reasonably accurate, the peak we observe in all the EDCs at $\Gamma$, located between the electron band bottom and $E_F$, must be the kagome flat band that has been brought down from above $E_F$.

Furthermore, considering that the FB is observable under all of the polarization geometries (see SM and Fig. S9 for the complete set of measurements), including those that only allow either even or odd parity orbitals to be observed, this suggests that both ${d_{xz}}$ and ${d_{yz}}$ orbitals are participating in forming the observed FB. 
%Specifically, in Fig.~\ref{fig:fig3}d and e, the observed FB can be exclusively assigned to having ${d_{xz}}$ and ${d_{yz}}$ orbital content, respectively, as ${d_{xy}}$ orbitals can not be observable at the first BZ center subjecting to photoemission matrix elements~\cite{Moser2017-jb}.
An additional confirmation for the observation of the FB below \ef~can be noted from the electron band at \g. This Sb ${p_{z}}$-dominated band is also seen in Cs\V, albeit with a shallower band bottom due to the smaller electron filling~\cite{Jeong2022-dd}. However, distinct from the Cs\V~case, we observe a bending of the electron dispersion near -70 meV in \Cr~(Fig.~\ref{fig:fig2}K).
%, as marked by a white arrow in Fig~\ref{fig:fig4} from the MDC-fitted band position. 
From the DFT calculations, it is clear that this electron band hybridizes with the kagome flat band where they cross (see SM and Fig. S9). Since the energy position of this slope change matches with the observed peak in the EDCs, this is an indication of the hybridization of the FB with the electron dispersion. We note that while electron-boson coupling generically could also result in dispersion kinks, in such situations a broadening of the imaginary part of the self-energy beyond the mode energy must also accompany the modification of the real part of the self-energy. Since we do not observe such broadening, we can definitively exclude electron-boson coupling as potential cause of such dispersion bending (see supplementary information S9 for detailed analysis). The observed hybridization of the flat band with the electron band further confirms that the FB we observe is intrinsic, not due to disorder-induced localized states that do not interact with the intrinsic band structure. The observation of this hybridization kink also confirms that the peak we observe in the EDC is not due to spectral weight leaking from a FB located above~\ef, but rather a FB that is located below \ef.

%revised by DJH
Having identified the kagome flat bands, we next present our RIXS results to reveal the low-energy excitations. While ARPES measures the single-particle spectral function in momentum space ($\bf k$), RIXS measures two-particle excitations in momentum transfer $ {\bf q} =\Delta {\bf k}$ such as magnetic excitations, phonons, crystal-field excitations and plasmons \cite{RevModPhys.83.705}. The nature of the excitations can be determined by polarization analysis. Specifically, magnetic excitations can be conclusively determined when the scattered photons have rotated polarization from the incident beam~\cite{Nozieres1974}. Such excitations have been observed in cuprates via cross polarization RIXS measurements~\cite{LeTacon2011, Dean2013, Jia_PRX_2016, Benjamin2014}.
Figure~\ref{fig:fig3}A illustrates the scattering geometry of our RIXS experiments. To conclusively determine spin excitations of the system and exclude other excitations, we used $\pi$-polarized light with 90$^\circ$ scattering angle between the incident and scattered X-rays of wave vectors ${\bf k}_{\rm in}$ and ${\bf k}_{\rm out}$, respectively. As the dominant low-energy electronic states of {\Cr} are derived from Cr 3$d$ orbitals, we carried out Cr $L_3$-edge RIXS to unravel its low-energy excitations. Figure~\ref{fig:fig3}B plots the X-ray absorption spectra across the Cr $L_3$-edge and shows X-ray energies at which RIXS measurements were taken. 
Figure~\ref{fig:fig3}C plots the RIXS spectra of {\Cr} taken at these different incident photon energies. The 90$^\circ$-scattering geometry with $\pi$-polarized X-rays ensures the measurement of spin excitations. A clear spectral feature of spin excitation appears around 70~meV as a shoulder of elastic scattering. This energy is smaller than the spin excitation energy scales of the iron~\cite{RevModPhys.87.855} and copper-oxide superconductors \cite{PhysRevLett.86.5377,RevModPhys.84.1383}.

To quantitatively determine the energy scale of the spin excitations in {\Cr}, we used a general damped harmonic oscillator model to generate the spectral profile of the spin excitations \cite{WOS:000316616400040,WOS:000798323000002} 
\begin{equation}
S(q_{||},E)=A\,\frac{E_0}{1-{\rm e}^{-\beta E}}\frac{2\,\gamma\,E}{\left(E^2-E_0^2\right)^2+(\gamma\, E)^2},
\label{Eq1}
\end{equation}
where $E_0$ is the undamped energy, $\gamma$ the damping factor, $\beta=\frac{1}{k_{\rm B}T}$ ($k_{\rm B}$ is the Boltzmann constant) and $A$ is a constant. The elastic peak was modeled using a Voigt function, with its spectral width determined by the instrumental resolution. The contribution from the fluorescence tail was described with a polynomial function. Since the spectral profile around zero energy was wider than the instrumental resolution, a low-energy component of spin excitation was required to fit the data. Therefore, the measured RIXS spectrum was fitted to a spectral profile comprising three components: one elastic and two electronic excitations, as shown in Fig.~\ref{fig:fig3}D. We note that we cannot describe the 
finite energy electronic excitations with a single peak because the overall fit would then require elastic scattering much larger than the instrumental resolution (see supplementary information S10).

We also measured momentum-dependent RIXS with $q_{\|}$ along ${\Gamma}-K$ (Fig.~\ref{fig:fig3}E) to further verify that the observed electronic excitations arise from transitions involving FBs. Figure~\ref{fig:fig3}F summarizes the $q_{\|}$ dependence of the magnetic excitations within the available momentum transfer range along the $[H,H,0]$ direction.  The fitted undamped energy $E_0$ of the two spin components in our RIXS data shows no dispersion for all $q_{\|}$ within the available range, as plotted in Fig.~\ref{fig:fig3}G. This is consistent with the FB results concluded from ARPES measurements.

Finally, we present the temperature dependence of the FBs and spin excitations across \tc. 
%To further determine if spin excitations seen in RIXS measurements are associated the FBs observed by ARPES experiments, we examine the temperature dependence of the FB across {\ts} (Fig.~\ref{fig:fig4}). 
In Fig.~\ref{fig:fig4}A we compare the band dispersion along $\bar{\Gamma}$-$\bar{M}$ taken with linear vertical (LV) polarization at 10 K (left) and 140 K (right). This experimental setup selects FB of ${d_{yz}}$ orbital (See SM Fig. S7). 
%From the EDC taken at \g, we can quantitatively extract the location of the FB by fitting the EDC, which gives a location of -76 meV as measured at 10 K (Fig.~\ref{fig:fig4}b-c).
The FBs are visible at both temperatures, as seen in the spectral image, the peak in the EDC, as well as from the bend in the fitted electron dispersion (marked by a white and red arrow respectively). From the spectral image, the FB location appears to be closer to \ef~at 140 K compared to 10 K. This can be seen better from the direct comparison of the EDC taken at the location marked by the black arrow: the peak is shifted towards \ef~at 140 K by about 20 meV, and is recovered after thermally cycling back to 10 K. The direction of the shift is opposite to that expected purely from the thermal broadening effect of the Fermi-Dirac distribution and hence indicates a real shift of the FB. Similarly, the comparison of band dispersion along $\bar{\Gamma}$-$\bar{M}$ taken with linear horizontal (LH) polarization and horizontal slit at 10 K (T $<$ {\tc}) and 140 K (T $>$ {\tc}) shows that FB of ${d_{xz}}$ orbital also shift towards {\ef} (Fig.~\ref{fig:fig4}E). We provide the continuous temperature evolution of this shift and the fitting result in Fig.~\ref{fig:fig4}F with the detailed temperature evolution presented in Fig.~\ref{fig:fig4}E. Additional temperature dependence measurements and data analysis are given in the supplementary information and Fig. S6.

We also conducted temperature-dependent RIXS measurements at temperatures across {\tc}. Figure~\ref{fig:fig4}G presents the measured RIXS spectra within an energy range of 0.3~eV.  All spectra were analyzed using the same curve-fitting scheme as shown in Fig.~\ref{fig:fig3}D. To highlight the spectral changes, Fig.~\ref{fig:fig4}H compares RIXS spectra above and below {\tc} after removing elastic scattering, revealing a subtle yet distinct broadening and shift towards higher energy as the temperature is cooled across \tc. For quantitative analysis, Figs. \ref{fig:fig4}I and J show the evolution of the fitted bare energy $E_0$ and damping factor $\gamma$ of both spin excitations, which align with ARPES results shown in Fig.~\ref{fig:fig4}F. 
The observed excitations are 
%understood to result from 
coupled to the electron-hole excitations across the Fermi level. As such,
the broadening and shifting of the spin excitations likely reflect the shifting of the FBs away from \ef, both below and above \ef.

%We also carried out temperature-dependent measurements of the RIXS spectra at the $q_\parallel$ marked in Fig.~\ref{fig:fig4}I, as shown in Fig.~\ref{fig:fig4}G. We have quantitatively extracted the damping factor $\gamma$, which appears to be constant above \tc~and increases below \tc. A similar trend is also observed for the energy, $E_0$, which increases below \tc. As the magnetic excitations are understood to arise from electron-hole excitations across the Fermi level, we can understand this widening of the excitations to be a reflection of the shifting of the FBs away from \ef~both below and above \ef, consistent with the observed shifting of FBs away from \ef~observed by our ARPES measurements. 

\section{Discussions}
First, from our systematic polarization dependence ARPES measurements, we clearly resolve the presence of the kagome flat bands near \ef~in \Cr. In comparison to the isostructural Cs\V, the Cr system is effectively electron-doped to an extent that at the DFT level, the kagome flat bands are brought much closer to \ef. However, our observations indicate that bare DFT does not accurately reproduce the flat bands below $E_F$. When the necessary correlation corrections are applied, the discrepancies in the flat band position can be reconciled at a qualitative level, although improved sample quality and refined theoretical approaches are encouraged to further resolve this matter \cite{Xie2024-ae, Wang2024-gn}.
This is reminiscent of the flat band observed in the 3D pyrochlore material CuV$_2$S$_4$, where the destructive interference and orbital-selective correlation effects work in tandem to pin the flat bands to \ef~\cite{Huang2024-aw}. It is also in line with the theoretical notion of emergent flat bands, the anticipated quantum phase transitions and strange metallicity~\cite{HuSciAdv2023,Chen_emergent_2024,Chen_arXiv_2023} and, by extension, unconventional superconductivity.

Second, since the temperature-dependent shift of the FBs away from the Fermi level observed by ARPES across \tc~is largely consistent with the increase in both the spin excitation energy and bandwidth observed by RIXS near the $\Gamma$ point, our results taken together provide compelling evidence that spin excitations probed by RIXS are coupled to the quasiparticle excitations of the FBs near the Fermi level. From previous X-ray diffraction experiments, it is clear that a structural phase transition occurs below \tc~in {\Cr}~\cite{Liu2023-vk}. Our RIXS results indicate that this transition is also associated with changes in spin excitations. Hence this transition is likely a composite order that involves both structural and magnetic degrees of freedom, potentially analogous to the structural and spin density wave order in the iron-pnictide superconductors, which calls for future neutron scattering experiments to elucidate.

Third, we discuss the potential mechanism for the involvement of the kagome FBs in the formation of this electronic order. 
%Electronically, a high density of states near the chemical potential is energetically unfavorable and could lead to electronic instabilities that result in its removal from the Fermi level, as is the case for Stoner-type ferromagnetism and VHS-driven nesting induced CDW order. 
Since the chemical potential of \Cr~happens to be near the kagome FBs, it is conceivable that their presence drives an electronic order that pushes the FBs away from the Fermi level, as we have observed. With hydrostatic pressure, this electronic order is suppressed, which would leave residual density of states from the FBs near \ef~to experience the quantum fluctuations expected near a quantum critical point and potentially enable superconductivity. Moreover, unlike the pinning of the large DOS to the M points of the BZ for the VHSs in a kagome metal, the kagome FB provides high density of states across a much larger portion of the BZ. Therefore, they could be susceptible to multiple types of electronic instabilities of similar energy scales and different $q$'s, which may be a cause for the involvement of both the lattice and spin degrees of freedom. It remains interesting to theoretically map out the competing orders promoted by the FBs that develop near-\ef~region in \Cr.

Finally, it is interesting to compare \Cr~to the other known kagome metal systems that have been extensively studied. First, A\V~and FeGe are both kagome metals that exhibit charge orders with an in-plane periodicity of $2\times2$. Both these systems exhibit the VHSs at the M points of the BZ in the proximity of \ef. While nesting is unlikely to be the dominant driving mechanism for the charge order as theoretically predicted, it may still be a necessary but insufficient condition for selecting the $q$ for the $2\times2$ charge order in these systems. In particular, A\V~has no magnetism from 3$d$ electrons but has coexisting superconductivity and charge order, while FeGe has a strong charge density wave and spin excitation coupling but without superconductivity \cite{teng2024spinchargelattice}.  {\Cr} appears to be special as suppression of the density wave by hydrostatic pressure drives strange metallicity and induces superconductivity \cite{Liu2023-vk}.

For kagome metals where the flat bands are in the vicinity of the Fermi level, there are two regimes where materials have been studied. Ni$_3$In represents a regime where the flat band is in proximity to \ef~yet not readily observed by photoemission. In this regime, no electronic orders are formed yet but the system exhibits non-Fermi liquid transport behavior indicating proximity to a potential quantum critical point~\cite{Ye2024-in}. Arguably in the opposite limit is the category of compounds that exhibit strong ordered magnetism. Kagome magnets including FeSn, FeGe, and the Mn- and Fe-based 166 systems all exhibit magnetism with ordering temperatures well above room temperature~\cite{Ye2018-iy, Teng2022-ru, Yin2020-ai, Mazet2002-sb}. The electronic structure of these compounds, when calculated for the paramagnetic state, all show kagome flat bands that live in the vicinity of the Fermi level, which in the magnetically ordered state split via the exchange splitting often with an energy scale of 1$\sim$2 eV~\cite{Teng2022-ru}. However, recent work on FeSn thin film that measures its electronic structure through its magnetic ordering temperature reveals that the exchange splitting of the bands remains largely intact above the magnetic ordering temperature, demonstrating that the origin of the magnetism in this system is local in nature (unpublished), which is likely common for these Fe- and Mn-based system with exceptionally high ordering temperatures. 

\Cr~is clearly different. Cr-based systems are typically magnetic, but more itinerant than Fe- and Mn-based systems. The similarity of the measured dispersions in the low-temperature density wave ordered phase to the DFT calculated band structure of the non-ordered phase indicates that the modification of the electronic structure through this order is subtle. This is not uncommon for systems with electronically driven orders with a similar ordering temperature, such as some of the underdoped iron-based superconductors where band folding due to the spin density wave is often hard to observe when close to the optimal doping~\cite{Ge2013-le, Liu2010-qr}. Yet, the flat band in \Cr~is clearly participating in the low temperature order, evident in its shift away from \ef~and its coupling with spin excitations. Hence \Cr~appears to exist in a regime that is also close to the potential quantum critical point of the phase diagram but on the ordered side~\cite{HuSciAdv2023,Chen_emergent_2024,Chen_arXiv_2023}, a place that is between the strong magnetically ordered kagome metals and Ni$_3$In. Recent theoretical studies have revealed that the flat bands in close proximity to \ef~in \Cr~give rise to antiferromagnetic fluctuations, suggesting the important role that these FBs play in this emergent order~\cite{Wu2024}. Interestingly, under the tuning knob of hydrostatic pressure, the competing phase in \Cr~is suppressed and superconductivity emerges. This drastic response to pressure, together with the observed FB near \ef, and its potential association with spin excitations suggest that \Cr~opens up access to a previously experimentally unexplored regime in the overarching phase diagram of kagome metals that offers intriguing insights into novel phases associated with the topological flat band physics.

\section{Methods}
\subsection{Crystal growth and characterization}
The {\Cr} single crystals were grown using the self-flux method. Cs (Solid, Alfa 99.8\%), Cr (Powder, Alfa 99.95\%), and Sb (Powder, Alfa 99.5\%) in a molar ratio of 12:3:30 were mixed. The mixture was loaded into an alumina crucible, and sealed in a Ta/Nb tube by arc welding under an argon atmosphere with one atmospheric pressure. The tube was sealed in an evacuated quartz tube to protect Ta/Nb from O${}_2$. The Ta/Nb tube was used to prevent the reaction between Cs vapor and the quartz tube, but Cs can still react slightly with Ta/Nb. The sample was heated to 850-905 ℃ within 12 h, kept for 50 h, cooled to 580-600 ℃ at a rate of 1.5-3 ℃/h, and cooled to room temperature naturally. Thin crystalline flakes can be found in the melts and the crystals are stable to water and the air. The sample size for this experiment is about 1 $\times$ 1 mm$^2$.

\subsection{ARPES measurements}
ARPES experiments were performed at the MAESTRO beamline of the Advanced Light Source and beamline 5-2 of the Stanford Synchrotron Radiation Lightsource.  
The MAESTRO beamline is equipped with a Scienta electron analyzer in a home-designed deflector mode and used a beamspot of $10\times10\,\mu\mathrm{m}^2$.  
SSRL beamline 5-2 employs a DA30 electron analyzer with a $10\times30\,\mu\mathrm{m}^2$ beamspot.  
The angular resolution was set to $0.1^\circ$ and the total energy resolution to $\leq20\,$meV.  
All samples were cleaved \emph{in situ} at $10\,$K, and measurements were conducted in ultra-high vacuum (base pressure $<5\times10^{-11}\,$Torr).  In figures, the error bars denote the standard errors $1\sigma$ in the fitted parameters, calculated as the square root of the diagonal elements of the covariance matrix returned by the least-squares fitting procedure.

\subsection{RIXS measurements}

We conducted Cr $L_3$-edge RIXS measurements using the AGM-AGS spectrometer at beamline 41A \cite{Singh2021} of the Taiwan Photon Source, National Synchrotron Radiation Research Center, Taiwan. The total RIXS energy resolution was 26 meV, determined by the spectral full width at half maximum of the elastic scattering with $\sigma$ polarization. The angle between the incident and scattered X-rays was fixed at $90^\circ$  for both incident-energy-dependent and temperature-dependent RIXS measurements, with unresolved polarization of scattered X-rays. Prior to XAS and RIXS measurements, crystallographic axes were aligned using hard X-ray diffraction with a specially designed tilting adjustment holder. Subsequently, samples were cleaved in air to expose a (001) surface. X-ray absorption spectra were acquired using a photodiode in fluorescence yield mode. The error bars of fitting constants represent the 1$\sigma$ standard errors on the fit parameters.

\subsection{DFT calculations}
All DFT calculations were performed with Vienna ab initio simulation package (VASP) code~\cite{Kresse1996a, Kresse1999},  with Perdew-Burke-Ernzerhof exchange-correlation functional~\cite{Perdew1996a}. The energy cutoff of plane wave basis is 450 eV, and 3D Brillouin zone is sampled with k-point mesh of $11\times11\times5$. All atoms are relaxed until residual force is under 0.01 eV/\AA. A tight-binding model of 31 orbitals is fitted from DFT results with Wannier functions, as implemented in Wannier90 package~\cite{pizzi2020}.

\subsection{XRD measurements}

The structure information of the crystal was investigated at ORNL using a Rigaku XtaLAB PRO diffractometer equipped with a HyPix-6000HE detector on single crystals with a dimension of $0.1 \times 0.1 \times 0.01$ mm$^3$. A molybdenum anode was used to generate x-rays with wavelength $\lambda$ = 0.7107 \AA. The samples were cooled by Helium gas flow provided by an Oxford N-Helix cryosystem.

\subsection{SEM and EDS measurements}

The chemcial composition is measured in the FEI Quanta 400 is a high resolution field emission scanning electron microscope. It is equipped with SE, backscatter, and EDS detectors, and can operate in high vac, low vac, and Wet modes. It is also equipped with a cooling stage. 

\section{Data Availability}
All data needed to evaluate the conclusions are present in the paper and supplementary materials. Additional data are available from the corresponding authors on request.

\section{Code Availability}
The band structure calculations and RIXS used in this study are available from the corresponding authors upon request.

\newpage
\section{References}
\bibliographystyle{naturemag}
\bibliography{Cr135.bib}

@article{Si23.1,
    author = {Si, Qimiao and Hussey, Nigel E.},
    title = "{Iron-based superconductors: Teenage, complex, challenging}",
    journal = {{Phys.\ Today}},
    volume = {76},
    number = {5},
    pages = {34-34},
    year = {2023},
    month = {05},
    doi = {10.1063/PT.3.5235},
    issn = {0031-9228}
    }

@article{Kresse1996a,
author = {Kresse, G. and Furthm{\"{u}}ller, J.},
doi = {10.1103/PhysRevB.54.11169},
issn = {0163-1829},
journal = {Phys. Rev. B},
month = {oct},
number = {16},
pages = {11169--11186},
title = {{Efficient iterative schemes for ab initio total-energy calculations using a plane-wave basis set}},
volume = {54},
year = {1996}
}

@article{Pizzi2020,
author = {Pizzi, Giovanni and Vitale, Valerio and Arita, Ryotaro and Bl{\"{u}}gel, Stefan and Freimuth, Frank and G{\'{e}}ranton, Guillaume and Gibertini, Marco and Gresch, Dominik and Johnson, Charles and Koretsune, Takashi and Iba{\~{n}}ez-Azpiroz, Julen and Lee, Hyungjun and Lihm, Jae-Mo and Marchand, Daniel and Marrazzo, Antimo and Mokrousov, Yuriy and Mustafa, Jamal I and Nohara, Yoshiro and Nomura, Yusuke and Paulatto, Lorenzo and Ponc{\'{e}}, Samuel and Ponweiser, Thomas and Qiao, Junfeng and Th{\"{o}}le, Florian and Tsirkin, Stepan S and Wierzbowska, Ma{\l}gorzata and Marzari, Nicola and Vanderbilt, David and Souza, Ivo and Mostofi, Arash A and Yates, Jonathan R},
doi = {10.1088/1361-648X/ab51ff},
issn = {0953-8984},
journal = {J. Phys.: Condensed Matter},
month = {apr},
number = {16},
pages = {165902},
title = {{Wannier90 as a community code: new features and applications}},
volume = {32},
year = {2020}
}

@article{Kresse1999,
author = {Kresse, G. and Joubert, D.},
doi = {10.1103/PhysRevB.59.1758},
issn = {0163-1829},
journal = {Phys. Rev. B},
month = {jan},
number = {3},
pages = {1758--1775},
title = {{From ultrasoft pseudopotentials to the projector augmented-wave method}},
volume = {59},
year = {1999}
}

@article{Perdew1996a,
author = {Perdew, John P. and Burke, Kieron and Ernzerhof, Matthias},
doi = {10.1103/PhysRevLett.77.3865},
issn = {0031-9007},
journal = {Phys. Rev. Lett.},
month = {oct},
number = {18},
pages = {3865--3868},
title = {{Generalized Gradient Approximation Made Simple}},
volume = {77},
year = {1996}
}

@ARTICLE{Wu2007-bs,
  title    = "Flat bands and Wigner crystallization in the honeycomb optical
              lattice",
  author   = "Wu, Congjun and Bergman, Doron and Balents, Leon and Das Sarma, S",
  journal  = "Phys. Rev. Lett.",
  volume   =  99,
  number   =  7,
  pages    = "070401",
  month    =  aug,
  year     =  2007,
  language = "en"
}

@ARTICLE{Mielke1991-hs,
  title     = "Ferromagnetic ground states for the Hubbard model on line graphs",
  author    = "Mielke, A",
  journal   = "J. Phys. A Math. Gen.",
  publisher = "IOP Publishing",
  volume    =  24,
  number    =  2,
  pages     = "L73",
  month     =  jan,
  year      =  1991,
  language  = "en"
}

@ARTICLE{Tasaki1992-ls,
  title    = "Ferromagnetism in the Hubbard models with degenerate
              single-electron ground states",
  author   = "Tasaki, H",
  journal  = "Phys. Rev. Lett.",
  volume   =  69,
  number   =  10,
  pages    = "1608--1611",
  month    =  sep,
  year     =  1992,
  language = "en"
}

@ARTICLE{Miyahara2007-dw,
  title   = "{BCS} theory on a flat band lattice",
  author  = "Miyahara, S and Kusuta, S and Furukawa, N",
  journal = "Physica C Supercond.",
  volume  = "460-462",
  pages   = "1145--1146",
  month   =  sep,
  year    =  2007
}

@ARTICLE{Zhang2010-qs,
  title     = "Proposed realization of itinerant ferromagnetism in optical
               lattices",
  author    = "Zhang, Shizhong and Hung, Hsiang-Hsuan and Wu, Congjun",
  journal   = "Phys. Rev. A",
  publisher = "American Physical Society",
  volume    =  82,
  number    =  5,
  pages     = "053618",
  month     =  nov,
  year      =  2010
}

@ARTICLE{Aoki2020-qm,
  title   = "Theoretical Possibilities for Flat Band Superconductivity",
  author  = "Aoki, Hideo",
  journal = "J. Supercond. Novel Magn.",
  volume  =  33,
  number  =  8,
  pages   = "2341--2346",
  month   =  aug,
  year    =  2020
}

@ARTICLE{Tang2011-jb,
  title    = "High-temperature fractional quantum Hall states",
  author   = "Tang, Evelyn and Mei, Jia-Wei and Wen, Xiao-Gang",
  journal  = "Phys. Rev. Lett.",
  volume   =  106,
  number   =  23,
  pages    = "236802",
  month    =  jun,
  year     =  2011,
  language = "en"
}

@ARTICLE{Bergman2008-xx,
  title     = "Band touching from real-space topology in frustrated hopping
               models",
  author    = "Bergman, Doron L and Wu, Congjun and Balents, Leon",
  journal   = "Phys. Rev. B Condens. Matter",
  publisher = "American Physical Society",
  volume    =  78,
  number    =  12,
  pages     = "125104",
  month     =  sep,
  year      =  2008
}

@ARTICLE{Kirchner2020-xc,
  title     = "Colloquium: Heavy-electron quantum criticality and
               single-particle spectroscopy",
  author    = "Kirchner, Stefan and Paschen, Silke and Chen, Qiuyun and Wirth,
               Steffen and Feng, Donglai and Thompson, Joe D and Si, Qimiao",
  journal   = "Rev. Mod. Phys.",
  publisher = "American Physical Society",
  volume    =  92,
  number    =  1,
  pages     = "011002",
  month     =  mar,
  year      =  2020
}

@ARTICLE{Sutherland1986-vu,
  title    = "Localization of electronic wave functions due to local topology",
  author   = "Sutherland, B",
  journal  = "Phys. Rev. B Condens. Matter",
  volume   =  34,
  number   =  8,
  pages    = "5208--5211",
  month    =  oct,
  year     =  1986,
  language = "en"
}

@ARTICLE{Yin2020-ai,
  title    = "Quantum-limit Chern topological magnetism in {TbMn$_6$Sn$_6$}",
  author   = "Yin, Jia-Xin and Ma, Wenlong and Cochran, Tyler A and Xu, Xitong
              and Zhang, Songtian S and Tien, Hung-Ju and Shumiya, Nana and
              Cheng, Guangming and Jiang, Kun and Lian, Biao and Song, Zhida
              and Chang, Guoqing and Belopolski, Ilya and Multer, Daniel and
              Litskevich, Maksim and Cheng, Zi-Jia and Yang, Xian P and
              Swidler, Bianca and Zhou, Huibin and Lin, Hsin and Neupert, Titus
              and Wang, Ziqiang and Yao, Nan and Chang, Tay-Rong and Jia,
              Shuang and Zahid Hasan, M",
  journal  = "Nature",
  volume   =  583,
  number   =  7817,
  pages    = "533--536",
  month    =  jul,
  year     =  2020,
  language = "en"
}

@ARTICLE{Kang2020-gy,
  title    = "Topological flat bands in frustrated kagome lattice {CoSn}",
  author   = "Kang, Mingu and Fang, Shiang and Ye, Linda and Po, Hoi Chun and
              Denlinger, Jonathan and Jozwiak, Chris and Bostwick, Aaron and
              Rotenberg, Eli and Kaxiras, Efthimios and Checkelsky, Joseph G
              and Comin, Riccardo",
  journal  = "Nat. Commun.",
  volume   =  11,
  number   =  1,
  pages    = "4004",
  month    =  aug,
  year     =  2020,
  language = "en"
}

@ARTICLE{Ye2018-iy,
  title     = "Massive Dirac fermions in a ferromagnetic kagome {metal:FeSn}",
  author    = "Ye, Linda and Kang, Mingu and Liu, Junwei and Von Cube, Felix
               and Wicker, Christina R and Suzuki, Takehito and Jozwiak, Chris
               and Bostwick, Aaron and Rotenberg, Eli and Bell, David C and Fu,
               Liang and Comin, Riccardo and Checkelsky, Joseph G",
  journal   = "Nature",
  publisher = "Nature Publishing Group",
  volume    =  555,
  number    =  7698,
  pages     = "638--642",
  year      =  2018
}

@ARTICLE{Peng2021-ig,
  title    = "Realizing Kagome Band Structure in {Two-Dimensional} Kagome
              Surface States of {RV$_6$Sn$_6$} ({R=Gd}, {Ho})",
  author   = "Peng, Shuting and Han, Yulei and Pokharel, Ganesh and Shen,
              Jianchang and Li, Zeyu and Hashimoto, Makoto and Lu, Donghui and
              Ortiz, Brenden R and Luo, Yang and Li, Houchen and Guo, Mingyao
              and Wang, Bingqian and Cui, Shengtao and Sun, Zhe and Qiao,
              Zhenhua and Wilson, Stephen D and He, Junfeng",
  journal  = "Phys. Rev. Lett.",
  volume   =  127,
  number   =  26,
  pages    = "266401",
  month    =  dec,
  year     =  2021,
  language = "en"
}

@ARTICLE{Kang2020-wq,
  title    = "Dirac fermions and flat bands in the ideal kagome metal {FeSn}",
  author   = "Kang, Mingu and Ye, Linda and Fang, Shiang and You, Jhih-Shih and
              Levitan, Abe and Han, Minyong and Facio, Jorge I and Jozwiak,
              Chris and Bostwick, Aaron and Rotenberg, Eli and Chan, Mun K and
              McDonald, Ross D and Graf, David and Kaznatcheev, Konstantine and
              Vescovo, Elio and Bell, David C and Kaxiras, Efthimios and van
              den Brink, Jeroen and Richter, Manuel and Prasad Ghimire, Madhav
              and Checkelsky, Joseph G and Comin, Riccardo",
  journal  = "Nat. Mater.",
  volume   =  19,
  number   =  2,
  pages    = "163--169",
  month    =  feb,
  year     =  2020,
  language = "en"
}

@ARTICLE{Lee2024-bc,
  title     = "Nature of charge density wave in kagome metal {ScV$_6$Sn$_6$}",
  author    = "Lee, Seongyong and Won, Choongjae and Kim, Jimin and Yoo,
               Jonggyu and Park, Sudong and Denlinger, Jonathan and Jozwiak,
               Chris and Bostwick, Aaron and Rotenberg, Eli and Comin, Riccardo
               and Kang, Mingu and Park, Jae-Hoon",
  journal   = "npj Quantum Materials",
  publisher = "Nature Publishing Group",
  volume    =  9,
  number    =  1,
  pages     = "15",
  month     =  jan,
  year      =  2024,
  language  = "en"
}

@ARTICLE{Kato2022-ow,
  title     = "Polarity-dependent charge density wave in the kagome
               superconductor {CsV$_{3}$Sb$_{5}$}",
  author    = "Kato, Takemi and Li, Yongkai and Nakayama, Kosuke and Wang,
               Zhiwei and Souma, Seigo and Kitamura, Miho and Horiba, Koji and
               Kumigashira, Hiroshi and Takahashi, Takashi and Sato, Takafumi",
  journal   = "Phys. Rev. B Condens. Matter",
  publisher = "American Physical Society",
  volume    =  106,
  number    =  12,
  pages     = "L121112",
  month     =  sep,
  year      =  2022
}

@ARTICLE{Hu2022-xa,
  title    = "Rich nature of Van Hove singularities in Kagome superconductor
              {CsV$_3$Sb$_5$}",
  author   = "Hu, Yong and Wu, Xianxin and Ortiz, Brenden R and Ju, Sailong and
              Han, Xinloong and Ma, Junzhang and Plumb, Nicholas C and Radovic,
              Milan and Thomale, Ronny and Wilson, Stephen D and Schnyder,
              Andreas P and Shi, Ming",
  journal  = "Nat. Commun.",
  volume   =  13,
  number   =  1,
  pages    = "2220",
  month    =  apr,
  year     =  2022,
  language = "en"
}

@ARTICLE{Checkelsky2024-zc,
  title     = "Flat bands, strange metals and the Kondo effect",
  author    = "Checkelsky, Joseph G and Bernevig, B Andrei and Coleman, Piers
               and Si, Qimiao and Paschen, Silke",
  journal   = "Nat. Rev. Mater.",
  publisher = "Nature Publishing Group",
  volume    = 9,
  pages     = "509-526",
  month     =  feb,
  year      =  2024,
  language  = "en"
}

@ARTICLE{Xie2021-fc,
  title    = "Fractional Chern insulators in magic-angle twisted bilayer
              graphene",
  author   = "Xie, Yonglong and Pierce, Andrew T and Park, Jeong Min and
              Parker, Daniel E and Khalaf, Eslam and Ledwith, Patrick and Cao,
              Yuan and Lee, Seung Hwan and Chen, Shaowen and Forrester, Patrick
              R and Watanabe, Kenji and Taniguchi, Takashi and Vishwanath,
              Ashvin and Jarillo-Herrero, Pablo and Yacoby, Amir",
  journal  = "Nature",
  volume   =  600,
  number   =  7889,
  pages    = "439--443",
  month    =  dec,
  year     =  2021,
  language = "en"
}

@ARTICLE{Das2021-ef,
  title     = "Symmetry-broken Chern insulators and Rashba-like Landau-level
               crossings in magic-angle bilayer graphene",
  author    = "Das, Ipsita and Lu, Xiaobo and Herzog-Arbeitman, Jonah and Song,
               Zhi-Da and Watanabe, Kenji and Taniguchi, Takashi and Bernevig,
               B Andrei and Efetov, Dmitri K",
  journal   = "Nat. Phys.",
  publisher = "Nature Publishing Group",
  volume    =  17,
  number    =  6,
  pages     = "710--714",
  month     =  mar,
  year      =  2021,
  language  = "en"
}

@ARTICLE{Lisi2020-jf,
  title     = "Observation of flat bands in twisted bilayer graphene",
  author    = "Lisi, Simone and Lu, Xiaobo and Benschop, Tjerk and de Jong,
               Tobias A and Stepanov, Petr and Duran, Jose R and Margot,
               Florian and Cucchi, Ir{\`e}ne and Cappelli, Edoardo and Hunter,
               Andrew and Tamai, Anna and Kandyba, Viktor and Giampietri,
               Alessio and Barinov, Alexei and Jobst, Johannes and Stalman,
               Vincent and Leeuwenhoek, Maarten and Watanabe, Kenji and
               Taniguchi, Takashi and Rademaker, Louk and van der Molen, Sense
               Jan and Allan, Milan P and Efetov, Dmitri K and Baumberger,
               Felix",
  journal   = "Nat. Phys.",
  publisher = "Nature Publishing Group",
  volume    =  17,
  number    =  2,
  pages     = "189--193",
  month     =  sep,
  year      =  2020,
  language  = "en"
}

@ARTICLE{Ortiz2021-zp,
  title     = "Superconductivity in the {Z$_{2}$} kagome metal
               {KV$_{3}$Sb$_{5}$}",
  author    = "Ortiz, Brenden R and Sarte, Paul M and Kenney, Eric M and Graf,
               Michael J and Teicher, Samuel M L and Seshadri, Ram and Wilson,
               Stephen D",
  journal   = "Phys. Rev. Mater.",
  publisher = "American Physical Society",
  volume    =  5,
  number    =  3,
  pages     = "034801",
  month     =  mar,
  year      =  2021
}

@ARTICLE{Ortiz2019-hh,
  title     = "New kagome prototype materials: discovery of
               {KV$_{3}$Sb$_{5}$}, {RbV$_{3}$Sb$_{5}$}, and {CsV$_{3}$Sb$_{5}$}",
  author    = "Ortiz, Brenden R and Gomes, L{\'\i}dia C and Morey, Jennifer R
               and Winiarski, Michal and Bordelon, Mitchell and Mangum, John S
               and Oswald, Iain W H and Rodriguez-Rivera, Jose A and Neilson,
               James R and Wilson, Stephen D and Ertekin, Elif and McQueen,
               Tyrel M and Toberer, Eric S",
  journal   = "Phys. Rev. Mater.",
  publisher = "American Physical Society",
  volume    =  3,
  number    =  9,
  pages     = "094407",
  month     =  sep,
  year      =  2019
}

@ARTICLE{Fernandes2022-tq,
  title    = "Iron pnictides and chalcogenides: a new paradigm for
              superconductivity",
  author   = "Fernandes, Rafael M and Coldea, Amalia I and Ding, Hong and
              Fisher, Ian R and Hirschfeld, P J and Kotliar, Gabriel",
  journal  = "Nature",
  volume   =  601,
  number   =  7891,
  pages    = "35--44",
  month    =  jan,
  year     =  2022,
  language = "en"
}

@ARTICLE{Armitage2010-px,
  title     = "Progress and perspectives on electron-doped cuprates",
  author    = "Armitage, N P and Fournier, P and Greene, R L",
  journal   = "Rev. Mod. Phys.",
  publisher = "American Physical Society",
  volume    =  82,
  number    =  3,
  pages     = "2421--2487",
  month     =  sep,
  year      =  2010
}

@ARTICLE{Zheng2022-qw,
  title    = "Emergent charge order in pressurized kagome superconductor
              {CsV$_3$Sb$_5$}",
  author   = "Zheng, Lixuan and Wu, Zhimian and Yang, Ye and Nie, Linpeng and
              Shan, Min and Sun, Kuanglv and Song, Dianwu and Yu, Fanghang and
              Li, Jian and Zhao, Dan and Li, Shunjiao and Kang, Baolei and
              Zhou, Yanbing and Liu, Kai and Xiang, Ziji and Ying, Jianjun and
              Wang, Zhenyu and Wu, Tao and Chen, Xianhui",
  journal  = "Nature",
  volume   =  611,
  number   =  7937,
  pages    = "682--687",
  month    =  nov,
  year     =  2022,
  language = "en"
}

@ARTICLE{Teng2023-ii,
  title     = "Magnetism and charge density wave order in kagome {FeGe}",
  author    = "Teng, Xiaokun and Oh, Ji Seop and Tan, Hengxin and Chen, Lebing
               and Huang, Jianwei and Gao, Bin and Yin, Jia-Xin and Chu,
               Jiun-Haw and Hashimoto, Makoto and Lu, Donghui and Jozwiak,
               Chris and Bostwick, Aaron and Rotenberg, Eli and Granroth,
               Garrett E and Yan, Binghai and Birgeneau, Robert J and Dai,
               Pengcheng and Yi, Ming",
  journal   = "Nat. Phys.",
  publisher = "Nature Publishing Group",
  volume    =  19,
  number    =  6,
  pages     = "814--822",
  month     =  mar,
  year      =  2023,
  language  = "en"
}

@ARTICLE{Teng2022-ru,
  title    = "Discovery of charge density wave in a kagome lattice
              antiferromagnet",
  author   = "Teng, Xiaokun and Chen, Lebing and Ye, Feng and Rosenberg,
              Elliott and Liu, Zhaoyu and Yin, Jia-Xin and Jiang, Yu-Xiao and
              Oh, Ji Seop and Hasan, M Zahid and Neubauer, Kelly J and Gao, Bin
              and Xie, Yaofeng and Hashimoto, Makoto and Lu, Donghui and
              Jozwiak, Chris and Bostwick, Aaron and Rotenberg, Eli and
              Birgeneau, Robert J and Chu, Jiun-Haw and Yi, Ming and Dai,
              Pengcheng",
  journal  = "Nature",
  volume   =  609,
  number   =  7927,
  pages    = "490--495",
  month    =  sep,
  year     =  2022,
  language = "en"
}

@ARTICLE{Huang2024-aw,
  title     = "{Non-Fermi} liquid behaviour in a correlated flat-band
               pyrochlore lattice",
  author    = "Huang, Jianwei and Chen, Lei and Huang, Yuefei and Setty,
               Chandan and Gao, Bin and Shi, Yue and Liu, Zhaoyu and Zhang,
               Yichen and Yilmaz, Turgut and Vescovo, Elio and Hashimoto,
               Makoto and Lu, Donghui and Yakobson, Boris I and Dai, Pengcheng
               and Chu, Jiun-Haw and Si, Qimiao and Yi, Ming",
  journal   = "Nat. Phys.",
  volume    = 20,
  publisher = "Nature Publishing Group",
  pages     = "603-609",
  month     =  jan,
  year      =  2024,
  language  = "en"
}

@ARTICLE{Liu2023-pr,
  title    = "Tunable Van Hove Singularity without Structural Instability in
              Kagome Metal {CsTi$_3$Bi$_5$}",
  author   = "Liu, Bo and Kuang, Min-Quan and Luo, Yang and Li, Yongkai and Hu,
              Cheng and Liu, Jiarui and Xiao, Qian and Zheng, Xiquan and Huai,
              Linwei and Peng, Shuting and Wei, Zhiyuan and Shen, Jianchang and
              Wang, Bingqian and Miao, Yu and Sun, Xiupeng and Ou, Zhipeng and
              Cui, Shengtao and Sun, Zhe and Hashimoto, Makoto and Lu, Donghui
              and Jozwiak, Chris and Bostwick, Aaron and Rotenberg, Eli and
              Moreschini, Luca and Lanzara, Alessandra and Wang, Yao and Peng,
              Yingying and Yao, Yugui and Wang, Zhiwei and He, Junfeng",
  journal  = "Phys. Rev. Lett.",
  volume   =  131,
  number   =  2,
  pages    = "026701",
  month    =  jul,
  year     =  2023,
  language = "en"
}

@ARTICLE{Rosenberg2022-do,
  title     = "Uniaxial ferromagnetism in the kagome metal {TbV$_{6}$Sn$_{6}$}",
  author    = "Rosenberg, Elliott and DeStefano, Jonathan M and Guo, Yucheng
               and Oh, Ji Seop and Hashimoto, Makoto and Lu, Donghui and
               Birgeneau, Robert J and Lee, Yongbin and Ke, Liqin and Yi, Ming
               and Chu, Jiun-Haw",
  journal   = "Phys. Rev. B Condens. Matter",
  publisher = "American Physical Society",
  volume    =  106,
  number    =  11,
  pages     = "115139",
  month     =  sep,
  year      =  2022
}

@ARTICLE{Liu2020-po,
  title    = "Orbital-selective Dirac fermions and extremely flat bands in
              frustrated kagome-lattice metal {CoSn}",
  author   = "Liu, Zhonghao and Li, Man and Wang, Qi and Wang, Guangwei and
              Wen, Chenhaoping and Jiang, Kun and Lu, Xiangle and Yan, Shichao
              and Huang, Yaobo and Shen, Dawei and Yin, Jia-Xin and Wang,
              Ziqiang and Yin, Zhiping and Lei, Hechang and Wang, Shancai",
  journal  = "Nat. Commun.",
  volume   =  11,
  number   =  1,
  pages    = "4002",
  month    =  aug,
  year     =  2020,
  language = "en"
}

@ARTICLE{Li2021-ll,
  title    = "Dirac cone, flat band and saddle point in kagome magnet {YMn$_6$Sn$_6$}",
  author   = "Li, Man and Wang, Qi and Wang, Guangwei and Yuan, Zhihong and
              Song, Wenhua and Lou, Rui and Liu, Zhengtai and Huang, Yaobo and
              Liu, Zhonghao and Lei, Hechang and Yin, Zhiping and Wang, Shancai",
  journal  = "Nat. Commun.",
  volume   =  12,
  number   =  1,
  pages    = "3129",
  month    =  may,
  year     =  2021,
  language = "en"
}

@ARTICLE{Lin2020-ns,
  title     = "Dirac fermions in antiferromagnetic {FeSn} kagome lattices with
               combined space inversion and time-reversal symmetry",
  author    = "Lin, Zhiyong and Wang, Chongze and Wang, Pengdong and Yi, Seho
               and Li, Lin and Zhang, Qiang and Wang, Yifan and Wang, Zhongyi
               and Huang, Hao and Sun, Yan and Huang, Yaobo and Shen, Dawei and
               Feng, Donglai and Sun, Zhe and Cho, Jun-Hyung and Zeng, Changgan
               and Zhang, Zhenyu",
  journal   = "Phys. Rev. B Condens. Matter",
  publisher = "American Physical Society",
  volume    =  102,
  number    =  15,
  pages     = "155103",
  month     =  oct,
  year      =  2020
}

@ARTICLE{Li2021-kp,
  title     = "Observation of Unconventional Charge Density Wave without
               Acoustic Phonon Anomaly in Kagome Superconductors
               {AV$_{3}$Sb$_{5}$} ({A=Rb}, {Cs})",
  author    = "Li, Haoxiang and Zhang, T T and Yilmaz, T and Pai, Y Y and
               Marvinney, C E and Said, A and Yin, Q W and Gong, C S and Tu, Z
               J and Vescovo, E and Nelson, C S and Moore, R G and Murakami, S
               and Lei, H C and Lee, H N and Lawrie, B J and Miao, H",
  journal   = "Phys. Rev. X",
  publisher = "American Physical Society",
  volume    =  11,
  number    =  3,
  pages     = "031050",
  month     =  sep,
  year      =  2021
}

@ARTICLE{Zhao2021-ii,
  title    = "Cascade of correlated electron states in the kagome
              superconductor {CsV$_3$Sb$_5$}",
  author   = "Zhao, He and Li, Hong and Ortiz, Brenden R and Teicher, Samuel M
              L and Park, Takamori and Ye, Mengxing and Wang, Ziqiang and
              Balents, Leon and Wilson, Stephen D and Zeljkovic, Ilija",
  journal  = "Nature",
  volume   =  599,
  number   =  7884,
  pages    = "216--221",
  month    =  nov,
  year     =  2021,
  language = "en"
}

@ARTICLE{Ye2024-in,
  title     = "Hopping frustration-induced flat band and strange metallicity in
               a kagome metal",
  author    = "Ye, Linda and Fang, Shiang and Kang, Mingu and Kaufmann, Josef
               and Lee, Yonghun and John, Caolan and Neves, Paul M and Zhao, S
               Y Frank and Denlinger, Jonathan and Jozwiak, Chris and Bostwick,
               Aaron and Rotenberg, Eli and Kaxiras, Efthimios and Bell, David
               C and Janson, Oleg and Comin, Riccardo and Checkelsky, Joseph G",
  journal   = "Nat. Phys.",
  publisher = "Nature Publishing Group",
  volume    = 20,
  pages     = "610-614",
  month     =  jan,
  year      =  2024,
  language  = "en"
}

@ARTICLE{Hu2023-za,
  title     = "Non-trivial band topology and orbital-selective electronic
               nematicity in a titanium-based kagome superconductor",
  author    = "Hu, Yong and Le, Congcong and Zhang, Yuhang and Zhao, Zhen and
               Liu, Jiali and Ma, Junzhang and Plumb, Nicholas C and Radovic,
               Milan and Chen, Hui and Schnyder, Andreas P and Wu, Xianxin and
               Dong, Xiaoli and Hu, Jiangping and Yang, Haitao and Gao,
               Hong-Jun and Shi, Ming",
  journal   = "Nat. Phys.",
  publisher = "Nature Publishing Group",
  volume    =  19,
  number    =  12,
  pages     = "1827--1833",
  month     =  sep,
  year      =  2023,
  language  = "en"
}

@ARTICLE{Cheng2023-hp,
  title    = "Visualization of Tunable Weyl Line in {A-A} Stacking Kagome
              Magnets",
  author   = "Cheng, Zi-Jia and Belopolski, Ilya and Tien, Hung-Ju and Cochran,
              Tyler A and Yang, Xian P and Ma, Wenlong and Yin, Jia-Xin and
              Chen, Dong and Zhang, Junyi and Jozwiak, Chris and Bostwick,
              Aaron and Rotenberg, Eli and Cheng, Guangming and Hossain, Md
              Shafayat and Zhang, Qi and Litskevich, Maksim and Jiang, Yu-Xiao
              and Yao, Nan and Schroeter, Niels B M and Strocov, Vladimir N and
              Lian, Biao and Felser, Claudia and Chang, Guoqing and Jia, Shuang
              and Chang, Tay-Rong and Hasan, M Zahid",
  journal  = "Adv. Mater.",
  volume   =  35,
  number   =  3,
  pages    = "e2205927",
  month    =  jan,
  year     =  2023,
  language = "en"
}

@ARTICLE{Cheng2024-vn,
  title     = "Nanoscale visualization and spectral fingerprints of the charge
               order in {ScV$_6$Sn$_6$} distinct from other kagome metals",
  author    = "Cheng, Siyu and Ren, Zheng and Li, Hong and Oh, Ji Seop and Tan,
               Hengxin and Pokharel, Ganesh and DeStefano, Jonathan M and
               Rosenberg, Elliott and Guo, Yucheng and Zhang, Yichen and Yue,
               Ziqin and Lee, Yongbin and Gorovikov, Sergey and Zonno, Marta
               and Hashimoto, Makoto and Lu, Donghui and Ke, Liqin and Mazzola,
               Federico and Kono, Junichiro and Birgeneau, R J and Chu,
               Jiun-Haw and Wilson, Stephen D and Wang, Ziqiang and Yan,
               Binghai and Yi, Ming and Zeljkovic, Ilija",
  journal   = "npj Quantum Materials",
  publisher = "Nature Publishing Group",
  volume    =  9,
  number    =  1,
  pages     = "14",
  month     =  jan,
  year      =  2024,
  language  = "en"
}

@ARTICLE{Hu2024-my,
  title    = "Phonon promoted charge density wave in topological kagome metal
              {ScV$_6$Sn$_6$}",
  author   = "Hu, Yong and Ma, Junzhang and Li, Yinxiang and Jiang, Yuxiao and
              Gawryluk, Dariusz Jakub and Hu, Tianchen and Teyssier,
              J{\'e}r{\'e}mie and Multian, Volodymyr and Yin, Zhouyi and Xu,
              Shuxiang and Shin, Soohyeon and Plokhikh, Igor and Han, Xinloong
              and Plumb, Nicholas C and Liu, Yang and Yin, Jia-Xin and
              Guguchia, Zurab and Zhao, Yue and Schnyder, Andreas P and Wu,
              Xianxin and Pomjakushina, Ekaterina and Hasan, M Zahid and Wang,
              Nanlin and Shi, Ming",
  journal  = "Nat. Commun.",
  volume   =  15,
  number   =  1,
  pages    = "1658",
  month    =  feb,
  year     =  2024,
  language = "en"
}

@ARTICLE{Jiang2021-we,
  title    = "Unconventional chiral charge order in kagome superconductor
              {KV$_3$Sb$_5$}",
  author   = "Jiang, Yu-Xiao and Yin, Jia-Xin and Denner, M Michael and
              Shumiya, Nana and Ortiz, Brenden R and Xu, Gang and Guguchia,
              Zurab and He, Junyi and Hossain, Md Shafayat and Liu, Xiaoxiong
              and Ruff, Jacob and Kautzsch, Linus and Zhang, Songtian S and
              Chang, Guoqing and Belopolski, Ilya and Zhang, Qi and Cochran,
              Tyler A and Multer, Daniel and Litskevich, Maksim and Cheng,
              Zi-Jia and Yang, Xian P and Wang, Ziqiang and Thomale, Ronny and
              Neupert, Titus and Wilson, Stephen D and Hasan, M Zahid",
  journal  = "Nat. Mater.",
  volume   =  20,
  number   =  10,
  pages    = "1353--1357",
  month    =  oct,
  year     =  2021,
  language = "en"
}

@ARTICLE{Yin2022-el,
  title    = "Discovery of Charge Order and Corresponding Edge State in Kagome
              Magnet {FeGe}",
  author   = "Yin, Jia-Xin and Jiang, Yu-Xiao and Teng, Xiaokun and Hossain, Md
              Shafayat and Mardanya, Sougata and Chang, Tay-Rong and Ye, Zijin
              and Xu, Gang and Denner, M Michael and Neupert, Titus and
              Lienhard, Benjamin and Deng, Han-Bin and Setty, Chandan and Si,
              Qimiao and Chang, Guoqing and Guguchia, Zurab and Gao, Bin and
              Shumiya, Nana and Zhang, Qi and Cochran, Tyler A and Multer,
              Daniel and Yi, Ming and Dai, Pengcheng and Hasan, M Zahid",
  journal  = "Phys. Rev. Lett.",
  volume   =  129,
  number   =  16,
  pages    = "166401",
  month    =  oct,
  year     =  2022,
  language = "en"
}

@ARTICLE{Liu2023-vk,
  title         = "Superconductivity under pressure in a chromium-based kagome metal",
  author        = "Liu, Yi and Liu, Zi-Yi and Bao, Jin-Ke and Yang, Peng-Tao
                   and Ji, Liang-Wen and Wu, Si-Qi and Shen, Qin-Xin and Luo,
                   Jun and Yang, Jie and Liu, Ji-Yong and Xu, Chen-Chao and
                   Yang, Wu-Zhang and Chai, Wan-Li and Lu, Jia-Yi and Liu,
                   Chang-Chao and Wang, Bo-Sen and Jiang, Hao and Tao, Qian and
                   Ren, Zhi and Xu, Xiao-Feng and Cao, Chao and Xu, Zhu-An and
                   Zhou, Rui and Cheng, Jin-Guang and Cao, Guang-Han",
  journal   = "Nature",
  volume   =  632,
  pages    = "1032",
  month    =  Aug,
  year     =  2024
}

@ARTICLE{Huang2023-he,
  title     = "Electron Correlations and Nematicity in the {Iron-Based}
               Superconductors",
  author    = "{Huang} and {Jianwei} and Guo, Yucheng and Yi, Ming",
  journal   = "Synchrotron Radiat. News",
  publisher = "Taylor \& Francis",
  volume    =  36,
  number    =  3,
  pages     = "30--38",
  month     =  may,
  year      =  2023
}

@ARTICLE{Huang2022-ih,
  title     = "Correlation-driven electronic reconstruction in {FeTe$_{1−x}$Se$_x$}",
  author    = "Huang, Jianwei and Yu, Rong and Xu, Zhijun and Zhu, Jian-Xin and
               Oh, Ji Seop and Jiang, Qianni and Wang, Meng and Wu, Han and
               Chen, Tong and Denlinger, Jonathan D and Mo, Sung-Kwan and
               Hashimoto, Makoto and Michiardi, Matteo and Pedersen, Tor M and
               Gorovikov, Sergey and Zhdanovich, Sergey and Damascelli, Andrea
               and Gu, Genda and Dai, Pengcheng and Chu, Jiun-Haw and Lu,
               Donghui and Si, Qimiao and Birgeneau, Robert J and Yi, Ming",
  journal   = "Communications Physics",
  publisher = "Nature Publishing Group",
  volume    =  5,
  number    =  1,
  pages     = "29",
  month     =  jan,
  year      =  2022,
  language  = "en"
}

@ARTICLE{Yi2013-bz,
  title    = "Observation of temperature-induced crossover to an
              orbital-selective Mott phase in {A$_x$Fe$_{2-y}$Se$_2$} ({A=K, Rb})
              superconductors",
  author   = "Yi, M and Lu, D H and Yu, R and Riggs, S C and Chu, J-H and Lv, B
              and Liu, Z K and Lu, M and Cui, Y-T and Hashimoto, M and Mo, S-K
              and Hussain, Z and Chu, C W and Fisher, I R and Si, Q and Shen,
              Z-X",
  journal  = "Phys. Rev. Lett.",
  volume   =  110,
  number   =  6,
  pages    = "067003",
  month    =  feb,
  year     =  2013,
  language = "en"
}

@ARTICLE{Yi2015-qj,
  title    = "Observation of universal strong orbital-dependent correlation
              effects in iron chalcogenides",
  author   = "Yi, M and Liu, Z-K and Zhang, Y and Yu, R and Zhu, J-X and Lee, J
              J and Moore, R G and Schmitt, F T and Li, W and Riggs, S C and
              Chu, J-H and Lv, B and Hu, J and Hashimoto, M and Mo, S-K and
              Hussain, Z and Mao, Z Q and Chu, C W and Fisher, I R and Si, Q
              and Shen, Z-X and Lu, D H",
  journal  = "Nat. Commun.",
  volume   =  6,
  pages    = "7777",
  month    =  jul,
  year     =  2015,
  language = "en"
}

@ARTICLE{Jeong2022-dd,
  title     = "Crucial role of out-of-plane Sb p orbitals in Van Hove
               singularity formation and electronic correlations in the
               superconducting kagome metal {CsV$_{3}$Sb$_{5}$}",
  author    = "Jeong, Min Yong and Yang, Hyeok-Jun and Kim, Hee Seung and Kim,
               Yong Baek and Lee, Sungbin and Han, Myung Joon",
  journal   = "Phys. Rev. B Condens. Matter",
  publisher = "American Physical Society",
  volume    =  105,
  number    =  23,
  pages     = "235145",
  month     =  jun,
  year      =  2022
}

@ARTICLE{Xie2024-ae,
  title         = "Electron correlations in the kagome flat band metal {CsCr$_3$Sb$_5$}",
  author        = "Xie, Fang and Fang, Yuan and Li, Ying and Huang, Yuefei and
                   Chen, Lei and Setty, Chandan and Sur, Shouvik and Yakobson,
                   Boris and Valent{\'\i}, Roser and Si, Qimiao",
  month         =  mar,
  year          =  2024,
  archivePrefix = "arXiv",
  journal   = "arXiv",
  primaryClass  = "cond-mat.str-el",
  eprint        = "2403.03911"
}

@ARTICLE{Mazet2002-sb,
  title   = "A study of the new {Yb$_{0.6}$Fe$_6$Sn$_6$} compound by neutron diffraction,
             {$^{57}$Fe} and {$^{119}$Sn} M{\"o}ssbauer spectroscopy experiments",
  author  = "Mazet, T and Isnard, O and Malaman, B",
  journal = "J. Magn. Magn. Mater.",
  volume  =  241,
  number  =  1,
  pages   = "51--59",
  month   =  mar,
  year    =  2002
}

@ARTICLE{Liu2010-qr,
  title   = "Evidence for a Lifshitz transition in electron-doped iron arsenic
             superconductors at the onset of superconductivity",
  author  = "Liu, Chang and Kondo, Takeshi and Fernandes, Rafael M and
             Palczewski, Ari D and Mun, Eun Deok and Ni, Ni and Thaler,
             Alexander N and Bostwick, Aaron and Rotenberg, Eli and Schmalian,
             J{\"o}rg and Bud'Ko, Sergey L and Canfield, Paul C and Kaminski,
             Adam",
  journal = "Nat. Phys.",
  volume  =  6,
  number  =  6,
  pages   = "419--423",
  year    =  2010
}

@ARTICLE{Ge2013-le,
  title     = "Anisotropic but Nodeless Superconducting Gap in the Presence of
               {Spin-Density} Wave in {Iron-Pnictide} Superconductor
               {NaFe$_{-x}$Co$_{x}$As}",
  author    = "Ge, Q Q and Ye, Z R and Xu, M and Zhang, Y and Jiang, J and Xie,
               B P and Song, Y and Zhang, C L and Dai, Pengcheng and Feng, D L",
  journal   = "Phys. Rev. X",
  publisher = "American Physical Society",
  volume    =  3,
  number    =  1,
  pages     = "011020",
  month     =  mar,
  year      =  2013
}

@ARTICLE{Xu2023-pn,
  title         = "Altermagnetic ground state in distorted Kagome metal {CsCr$_3$Sb$_5$}",
  author        = "Xu, Chenchao and Wu, Siqi and Zhi, Guo-Xiang and Cao, Guanghan and Dai, Jianhui and Cao, Chao and Wang, Xiaoqun and Lin, Hai-Qing",
  Journal = {Nat. Commun.},
  Year = {2025},
  Volume = {16},
  Month = {april},
DOI = {10.1038/s41467-025-58446-6},
pages = {3114}
}

@ARTICLE{Wang2024-gn,
  title = {Heavy fermions in frustrated Hund's metal with portions of incipient flat bands},
  author = {Wang, Yilin},
  journal = {Phys. Rev. B},
  volume = {111},
  issue = {3},
  pages = {035127},
  numpages = {8},
  year = {2025},
  month = {Jan},
  publisher = {American Physical Society},
  doi = {10.1103/PhysRevB.111.035127},
  url = {https://link.aps.org/doi/10.1103/PhysRevB.111.035127}
}

@ARTICLE{Wu2024,
  title         = "Flat-band enhanced antiferromagnetic fluctuations and superconductivity in pressurized {CsCr$_3$Sb$_5$}",
  author        = "Wu, Siqi and Xu, Chenchao and Wang, Xiaoqun and Lin, Hai-Qing and Cao, Chao and Cao, Guang-Han",
  Journal = {Nat. Commun.},
  Year = {2025},
  Volume = {16},
  Month = {Jan},
DOI = {10.1038/s41467-025-56582-7},
pages = {1375},
}

@article{RevModPhys.83.705,
  title = {Resonant inelastic x-ray scattering studies of elementary excitations},
  author = {Ament, Luuk J. P. and van Veenendaal, Michel and Devereaux, Thomas P. and Hill, John P. and van den Brink, Jeroen},
  journal = {Rev. Mod. Phys.},
  volume = {83},
  issue = {2},
  pages = {705--767},
  numpages = {0},
  year = {2011},
  month = {Jun},
  publisher = {American Physical Society},
  doi = {10.1103/RevModPhys.83.705},
}

@article{RevModPhys.87.855,
  title = {Antiferromagnetic order and spin dynamics in iron-based superconductors},
  author = {Dai, Pengcheng},
  journal = {Rev. Mod. Phys.},
  volume = {87},
  issue = {3},
  pages = {855--896},
  numpages = {42},
  year = {2015},
  month = {Aug},
  publisher = {American Physical Society},
  doi = {10.1103/RevModPhys.87.855},
}

@article{PhysRevLett.86.5377,
  title = {Spin Waves and Electronic Interactions in {La$_2$CuO$_4$}},
  author = {Coldea, R. and Hayden, S. M. and Aeppli, G. and Perring, T. G. and Frost, C. D. and Mason, T. E. and Cheong, S.-W. and Fisk, Z.},
  journal = {Phys. Rev. Lett.},
  volume = {86},
  issue = {23},
  pages = {5377--5380},
  numpages = {0},
  year = {2001},
  month = {Jun},
  publisher = {American Physical Society},
  doi = {10.1103/PhysRevLett.86.5377},
}

@article{RevModPhys.84.1383,
  title = {A common thread: The pairing interaction for unconventional superconductors},
  author = {Scalapino, D. J.},
  journal = {Rev. Mod. Phys.},
  volume = {84},
  issue = {4},
  pages = {1383--1417},
  numpages = {0},
  year = {2012},
  month = {Oct},
  publisher = {American Physical Society},
  doi = {10.1103/RevModPhys.84.1383},
}

@article{WOS:000316616400040,
Author = {Zhou, Ke-Jin and Huang, Yao-Bo and Monney, Claude and Dai, Xi and
   Strocov, Vladimir N. and Wang, Nan-Lin and Chen, Zhi-Guo and Zhang,
   Chenglin and Dai, Pengcheng and Patthey, Luc and van den Brink, Jeroen
   and Ding, Hong and Schmitt, Thorsten},
Title = {Persistent high-energy spin excitations in iron-pnictide superconductors},
Journal = {Nat. Commun.},
Year = {2013},
Volume = {4},
Month = {FEB},
DOI = {10.1038/ncomms2428},
pages = {1470},
ISSN = {2041-1723},
ResearcherID-Numbers = {huang, yao/GWR-5388-2022
   Zhang, Cheng/GRS-8698-2022
   van den Brink, Jeroen/Y-3931-2019
   van den Brink, Jeroen/E-5670-2011
   Patthey, Luc/G-6130-2018
   Wang, Nan/HLV-7836-2023
   Wang, Nan/GRY-3150-2022
   Dai, Xi/C-4236-2008
   Chen, Zhiguo/B-9192-2015
   Dai, Xi/HCG-9886-2022
   Zhang, Cheng/JAC-5078-2023
   Schmitt, Thorsten/A-7025-2010
   Dai, Pengcheng/C-9171-2012
   Monney, Claude/C-5553-2011
   },
ORCID-Numbers = {van den Brink, Jeroen/0000-0001-6594-9610
   Patthey, Luc/0000-0001-6101-8069
   Dai, Xi/0000-0002-2396-0966
   Chen, Zhiguo/0000-0002-8242-4784
   Zhang, Cheng/0000-0003-3537-0434
   Dai, Pengcheng/0000-0002-6088-3170
   Monney, Claude/0000-0003-3496-1435
   Zhou, Ke-Jin/0000-0001-9293-0595},
Unique-ID = {WOS:000316616400040},
}

@article{WOS:000798323000002,
Author = {Lu, Xingye and Zhang, Wenliang and Tseng, Yi and Liu, Ruixian and Tao,
   Zhen and Paris, Eugenio and Liu, Panpan and Chen, Tong and Strocov,
   Vladimir N. and Song, Yu and Yu, Rong and Si, Qimiao and Dai, Pengcheng
   and Schmitt, Thorsten},
Title = {Spin-excitation anisotropy in the nematic state of detwinned FeSe},
Journal = {Nat. Phys.},
Year = {2022},
Volume = {18},
Number = {7},
Pages = {806+},
Month = {JUL},
DOI = {10.1038/s41567-022-01603-1},
EarlyAccessDate = {MAY 2022},
ISSN = {1745-2473},
EISSN = {1745-2481},
ResearcherID-Numbers = {Yu, Song/JMR-3167-2023
   Schmitt, Thorsten/A-7025-2010
   song, yu/KCZ-2003-2024
   Zhang, Wenliang/HZI-6186-2023
   liu, pan/HIR-9103-2022
   Lu, Xingye/Q-4184-2019
   Dai, Pengcheng/C-9171-2012},
ORCID-Numbers = {Zhang, Wenliang/0000-0003-3278-4076
   Lu, Xingye/0000-0002-0409-1240
   Tseng, Yi/0000-0001-6788-4398
   Schmitt, Thorsten/0000-0002-5737-1094
   Song, Yu/0000-0002-8835-9071
   Dai, Pengcheng/0000-0002-6088-3170},
Unique-ID = {WOS:000798323000002},
}

@article{teng2024spinchargelattice,
  title = {Spin-Charge-Lattice Coupling across the Charge Density Wave Transition in a Kagome Lattice Antiferromagnet},
  author = {Teng, Xiaokun and Tam, David W. and Chen, Lebing and Tan, Hengxin and Xie, Yaofeng and Gao, Bin and Granroth, Garrett E. and Ivanov, Alexandre and Bourges, Philippe and Yan, Binghai and Yi, Ming and Dai, Pengcheng},
  journal = {Phys. Rev. Lett.},
  volume = {133},
  issue = {4},
  pages = {046502},
  numpages = {7},
  year = {2024},
  month = {Jul},
  publisher = {American Physical Society},
  doi = {10.1103/PhysRevLett.133.046502},
  url = {https://link.aps.org/doi/10.1103/PhysRevLett.133.046502}
}

@article{Singh2021,
  title={Development of the Soft {X-ray} {AGM-AGS RIXS} beamline at the Taiwan Photon Source},
  author={Singh, A and Huang, H. Y. and Chu, Y. Y. and Hua, C. Y. and Lin, S. W. and Fung, H. S. and Shiu, H. W. and Chang, J. and Li, J. H. and Okamoto, J and  Chiu, C. C. and Chang, C. H. and Wu, W. B. and Perng, S. Y. and Chung, S. C. and Kao, K. Y. and Yeh, S. C. and Chao, H. Y. and Chen, J. H. and Huang, D. J. and C. T. Chen},
  journal={J. Synchrotron Radiat.},
  volume={28},
  pages={977},
  year={2021},
  publisher={International Union of Crystallography}
}

@article{HuSciAdv2023,
author = {Haoyu Hu  and Qimiao Si },
title = {Coupled topological flat and wide bands: Quasiparticle formation and destruction},
journal = {Sci. Advances},
volume = {9},
number = {29},
pages = {eadg0028},
year = {2023},
doi = {10.1126/sciadv.adg0028}}

@article{Chen_emergent_2024,
	title = {Emergent flat band and topological Kondo semimetal driven by orbital-selective correlations},
	volume = {15},
	issn = {2041-1723},
	doi = {10.1038/s41467-024-49306-w},
	number = {1},
	journal = {Nat. Commun.},
	author = {Chen, Lei and Xie, Fang and Sur, Shouvik and Hu, Haoyu and Paschen, Silke and Cano, Jennifer and Si, Qimiao},
	month = jun,
	year = {2024},
	pages = {5242}
}

@misc{Chen_arXiv_2023,
      title={Metallic quantum criticality enabled by flat bands in a kagome lattice}, 
      author={Lei Chen and Fang Xie and Shouvik Sur and Haoyu Hu and Silke Paschen and Jennifer Cano and Qimiao Si},
      year={2023},
      eprint={2307.09431},
      archivePrefix={arXiv},
      journal   = {"arXiv"}
}

@article{Werhahn2022,
title = {The kagomé metals {RbTi$_3$Bi$_5$} and {CsTi$_3$Bi$_5$}},
author = {Dominik Werhahn and Brenden R. Ortiz and Aurland K. Hay and Stephen D. Wilson and Ram Seshadri and Dirk Johrendt},
pages = {757--764},
volume = {77},
number = {11-12},
journal = {Zeitschrift für Naturforschung B},
doi = {doi:10.1515/znb-2022-0125},
year = {2022}
}

@misc{Yang_arXiv_2022,
      title={Titanium-based kagome superconductor {CsTi$_3$Bi$_5$} and topological states}, 
      author={Haitao Yang and Zhen Zhao and Xin-Wei Yi and Jiali Liu and Jing-Yang You and Yuhang Zhang and Hui Guo and Xiao Lin and Chengmin Shen and Hui Chen and Xiaoli Dong and Gang Su and Hong-Jun Gao},
      year={2022},
      eprint={2209.03840},
      archivePrefix={arXiv},
      journal   = {"arXiv"}
}

@article{Samanta2024,
  title={Emergence of flat bands and ferromagnetic fluctuations via orbital-selective electron correlations in Mn-based kagome metal},
  author={Subhasis Samanta and Hwiwoo Park and Chanhyeon Lee and Sungmin Jeon and Hengbo Cui and Yong-Xin Yao and Jungseek Hwang and Kwang-Yong Choi and Heung-Sik Kim},
  journal={Nat. Comm.},
  volume={15},
  pages={5376},
  year={2024}
}

@article{Lieb1989,
  title = {Two theorems on the Hubbard model},
  author = {Lieb, Elliott H.},
  journal = {Phys. Rev. Lett.},
  volume = {62},
  issue = {10},
  pages = {1201--1204},
  numpages = {0},
  year = {1989},
  month = {Mar},
  publisher = {American Physical Society},
  doi = {10.1103/PhysRevLett.62.1201},
}

@Article{Calugaru2022,
author={C{\u{a}}lug{\u{a}}ru, Dumitru
and Chew, Aaron
and Elcoro, Luis
and Xu, Yuanfeng
and Regnault, Nicolas
and Song, Zhi-Da
and Bernevig, B. Andrei},
title={General construction and topological classification of crystalline flat bands},
journal={Nat. Phys.},
year={2022},
month={Feb},
day={01},
volume={18},
number={2},
pages={185-189},
issn={1745-2481},
doi={10.1038/s41567-021-01445-3},
}

@Article{Regnault2022,
author={Regnault, Nicolas
and Xu, Yuanfeng
and Li, Ming-Rui
and Ma, Da-Shuai
and Jovanovic, Milena
and Yazdani, Ali
and Parkin, Stuart S. P.
and Felser, Claudia
and Schoop, Leslie M.
and Ong, N. Phuan
and Cava, Robert J.
and Elcoro, Luis
and Song, Zhi-Da
and Bernevig, B. Andrei},
title={Catalogue of flat-band stoichiometric materials},
journal={Nature},
year={2022},
month={Mar},
day={01},
volume={603},
number={7903},
pages={824-828},
issn={1476-4687},
doi={10.1038/s41586-022-04519-1},
}

@article{Nozieres1974,
  title = {Threshold singularities of the x-ray Raman scattering in metals},
  author = {Nozi\`eres, Philippe and Abrahams, Elihu},
  journal = {Phys. Rev. B},
  volume = {10},
  issue = {8},
  pages = {3099--3112},
  numpages = {0},
  year = {1974},
  month = {Oct},
  publisher = {American Physical Society},
  doi = {10.1103/PhysRevB.10.3099},
}

@article{Benjamin2014,
  title = {Single-Band Model of Resonant Inelastic X-Ray Scattering by Quasiparticles in High-${T}_{c}$ Cuprate Superconductors},
  author = {Benjamin, David and Klich, Israel and Demler, Eugene},
  journal = {Phys. Rev. Lett.},
  volume = {112},
  issue = {24},
  pages = {247002},
  numpages = {5},
  year = {2014},
  month = {Jun},
  publisher = {American Physical Society},
  doi = {10.1103/PhysRevLett.112.247002},
}

@article{Jia_PRX_2016,
  title = {Using RIXS to Uncover Elementary Charge and Spin Excitations},
  author = {Jia, Chunjing and Wohlfeld, Krzysztof and Wang, Yao and Moritz, Brian and Devereaux, Thomas P.},
  journal = {Phys. Rev. X},
  volume = {6},
  issue = {2},
  pages = {021020},
  numpages = {16},
  year = {2016},
  month = {May},
  publisher = {American Physical Society},
  doi = {10.1103/PhysRevX.6.021020},
}

@Article{LeTacon2011,
author={M Le Tacon and G Ghiringhelli and J Chaloupka and M Moretti Sala and V Hinkov and MW Haverkort and M Minola and M Makr and KJ Zhou and S Blanco-Canosa and C Monney and YT Song and GL Sun and CT Lin and GM De Luca and M Salluzzo and G Khaliullin and T Schmitt and L Braicovich and B Keimer},
title={Intense paramagnon excitations in a large family of high-temperature superconductors},
journal={Nat.Phys.},
year={2011},
volume={7},
pages={725-730}
}

@Article{Dean2013,
author={MPM Dean and G Dellea and RS Springell and F Yakhou-Harris and K Kummer and NB Brookes and X Liu and Y-J Sun and J Strle and T Schmitt and L Braicovich and G Ghiringhelli and I Bozovic and JP Hill},
title={Persistence of magnetic excitations in {La$_{2−x}$Sr$_x$CuO$_4$} from the undoped insulator to the heavily overdoped non-superconducting metal},
journal={Nat. Materials},
year={2013},
volume={12},
pages={1019-1023}
}

@Article{Essafi2017,
author={Essafi, K. and Jaubert, L. D.~C. and Udagawa, M.},
title={Flat bands and Dirac cones in breathing lattices},
journal={J. Physics: Condensed Matter},
year={2017},
volume={29},
pages={315802}
}

@article{VENTURINI199135,
title = {Magnetic properties of {RMn$_6$Sn$_6$} ({R} = {Sc}, {Y}, {Gd−Tm}, {Lu}) compounds with {HfFe$_6$Ge$_6$} type structure},
journal = {J. Magnetism and Magnetic Materials},
volume = {94},
number = {1},
pages = {35-42},
year = {1991},
issn = {0304-8853},
doi = {https://doi.org/10.1016/0304-8853(91)90108-M},
author = {G. Venturini and B.Chafik El Idrissi and B. Malaman}
}

@article{Neupane2009,
  title = {Observation of a Novel Orbital Selective Mott Transition in {Ca$_{1.8}$Sr$_{0.2}$RuO$_4$}},
  author = {Neupane, M. and Richard, P. and Pan, Z.-H. and Xu, Y.-M. and Jin, R. and Mandrus, D. and Dai, X. and Fang, Z. and Wang, Z. and Ding, H.},
  journal = {Phys. Rev. Lett.},
  volume = {103},
  issue = {9},
  pages = {097001},
  numpages = {4},
  year = {2009},
  month = {Aug},
  publisher = {American Physical Society},
  doi = {10.1103/PhysRevLett.103.097001},
}

@ARTICLE{Tian2023-ed,
  title     = "Evidence for Dirac flat band superconductivity enabled by quantum
               geometry",
  author    = "Tian, Haidong and Gao, Xueshi and Zhang, Yuxin and Che, Shi and
               Xu, Tianyi and Cheung, Patrick and Watanabe, Kenji and Taniguchi,
               Takashi and Randeria, Mohit and Zhang, Fan and Lau, Chun Ning and
               Bockrath, Marc W",
  journal   = "Nature",
  publisher = "Springer Science and Business Media LLC",
  volume    =  614,
  number    =  7948,
  pages     = "440--444",
  month     =  feb,
  year      =  2023,
  language  = "en"
}

@ARTICLE{Iglovikov2014-jd,
  title     = "Superconducting transitions in flat-band systems",
  author    = "Iglovikov, V I and Hébert, F and Grémaud, B and Batrouni, G G and
               Scalettar, R T",
  journal   = "Phys. Rev. B",
  publisher = "American Physical Society",
  volume    =  90,
  number    =  9,
  pages     =  094506,
  month     =  sep,
  year      =  2014,
  language  = "en"
}

@ARTICLE{Bistritzer2011-qh,
  title     = "Moiré bands in twisted double-layer graphene",
  author    = "Bistritzer, Rafi and MacDonald, Allan H",
  journal   = "Proc. Natl. Acad. Sci. U. S. A.",
  publisher = "Proceedings of the National Academy of Sciences",
  volume    =  108,
  number    =  30,
  pages     = "12233--12237",
  month     =  jul,
  year      =  2011,
  language  = "en"
}

@ARTICLE{Cao2018,
  title     = "Unconventional superconductivity in magic-angle graphene superlattices",
  author    = "Cao, Y. and Fatemi, V. and Fang, S. and Watanabe, K. and Taniguchi, T. and Kaxiras, E. and Jarillo-Herrero, P.",
  journal   = "Nature",
  volume    =  556,
  pages     = "43--50",
  year      =  2018,
  language  = "en"
}

\newpage

\section{Acknowledgments}
The single-crystal synthesis work and RIXS experiments at Rice were supported by the U.S. DOE, BES under Grant No. DE-SC0012311 (P.D.). Part of the materials characterization efforts at Rice is supported by the Robert A. Welch Foundation Grant No. C-1839 (P.D.).  
The ARPES work at Rice University was supported by the U.S. DOE grant No. DE-SC0021421, the Gordon and Betty Moore Foundation's EPiQS Initiative through grant No. GBMF9470 and the Robert A. Welch Foundation Grant No. C-2175 (M.Y.). 
Y.G. is supported in part by an ALS Doctoral Fellowship in Residence.
Y.Z. is partially supported by the Air Force Office of Scientific Research (AFOSR) Grant No. FA9550-21-1-0343. 
The RIXS work in Taiwan is partially supported by the National Science and Technology Council of Taiwan under Grant No. NSTC 112-2112-M-213-026-MY3 (D.J.H.)
The theory work at Rice is supported 
by the NSF Grant No. DMR-2220603
(F.X. and Y.F.),  by the AFOSR Grant No. FA9550-21-1-0356 (Y.W.), and by the Robert A. Welch Foundation Grant No. C-1411 (Q.S.), and by the Vannevar Bush Faculty Fellowship ONR-VB N00014-23-1-2870 (Q.S.). 
Computational modeling was supported by the Office of Naval Research Grant N00014-22-1-2753 (Y.H. and B.I.Y.).
The transport and thermodynamic measurements at UW were supported by the Air Force Office of Scientific Research (AFOSR) under Award No. FA2386-21-1-4060 and the David Lucile Packard Foundation (J.H.C).
Work at the University of California, Berkeley and Lawrence Berkeley National Laboratory was funded by the U.S. DOE, Office of Science, Office of Basic Energy Sciences, Materials Sciences and Engineering Division under Contract No. DE-AC02-05CH11231 (Quantum Materials Program KC2202). J.S.O, R.J.B., and M.Y. acknowledge the support from National Science Foundation (NSF) grants Nos. DMR-1921798 and DMR-2324032.
M.H. and D.L. acknowledge the support of the U.S. Department of Energy, Office of Science, Office of Basic Energy Sciences, Division of Material Sciences and Engineering, under contract DE-AC02-76SF00515.
A.F. acknowledges the support of the National Science and Technology Council of Taiwan under Grant No. 113-2112-M-007-033, the Japan Society for the Promotion of Science under Grant No. JP22K03535, and the Yushan Fellow Program of the Ministry of Education of Taiwan.
This research used resources of the Advanced Light Source, which is a DOE Office of Science User Facility under contract no. DE-AC02-05CH11231. Use of the Stanford Synchrotron Radiation Lightsource, SLAC National Accelerator Laboratory, is supported by the U.S. Department of Energy, Office of Science, Office of Basic Energy Sciences under Contract No. DE-AC02-76SF00515. A portion of this research used resources at the Spallation Neutron Source, a DOE Office of Science User Facility operated by Oak Ridge National Laboratory. 

\section{Author contributions}
PD initiated this project. PD, MY, DJH, and QS oversaw the project. YG, JSO, ZR, YZ, ZY, AB (Ananya Biswas) and CH carried out the ARPES measurements with the help of ER, AB (Aaron Bostwick), CJ, MH, DL,  JK and RJB. The ARPES data were analyzed by YG and HW with the help of MY. HYH, JO, and GC conducted the RIXS measurements with the help of DJH, AF, and CTC.  HYH and DJH analyzed the RIXS data with the help of ZW and XL. Single crystals were synthesized by ZW and BG under the guidance of PD and GHC. $U$(1) auxiliary-spin calculations were carried out by FX, YF, YW and QS. Density-functional theory calculations and tight-binding model fitting were carried out by YH under the guidance of BY. Transport measurements were carried out by ZL and JC. FY carried out X-ray diffraction measurements. YG, ZW, MY, DJH, and PD wrote the paper with input from all co-authors.

\section{Competing interests}
The authors declare no competing interests. 

\newpage

\begin{figure*}
\includegraphics[width=0.99\textwidth]{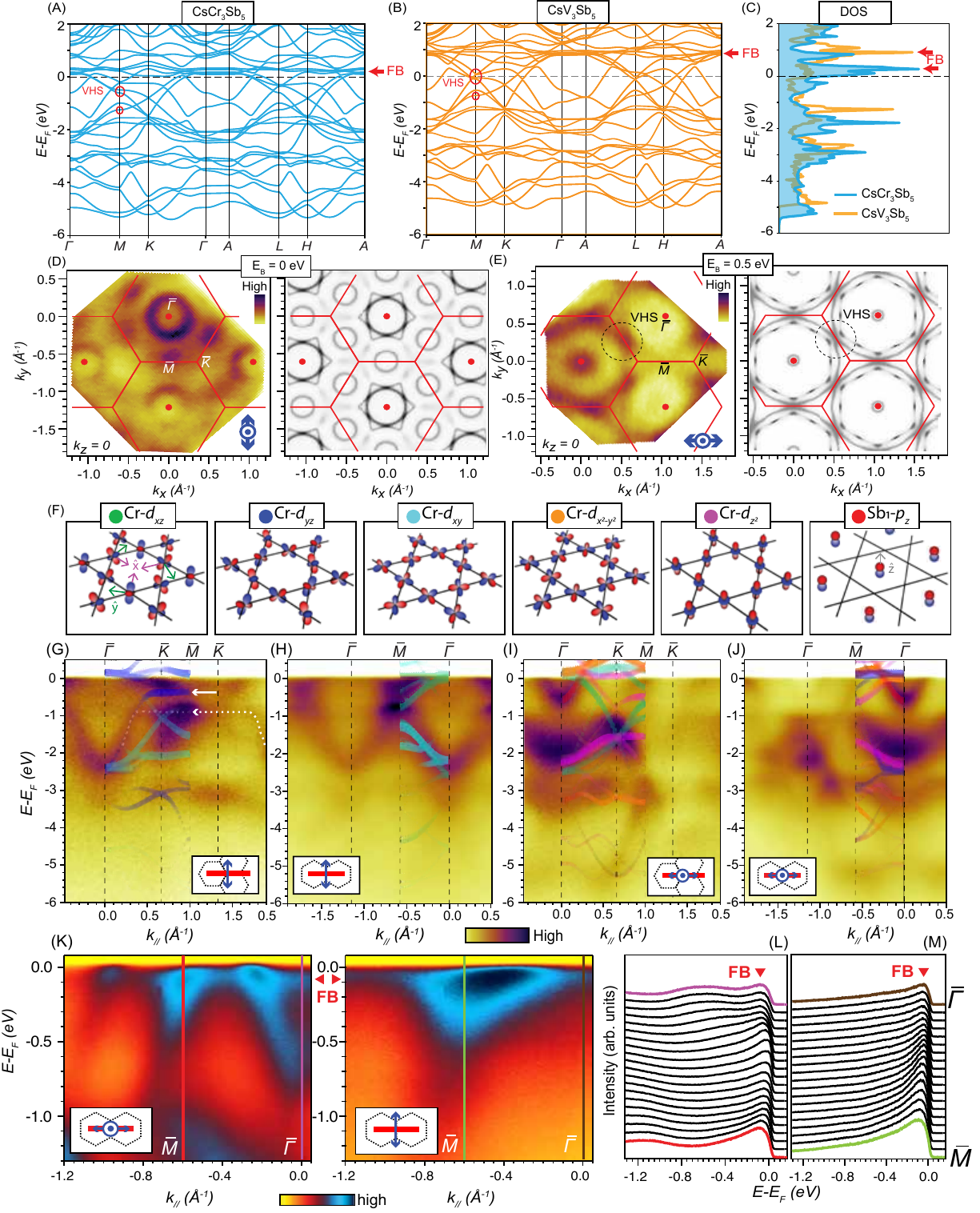}
\caption{\label{fig:fig2} \textbf{Electronic structure of {\Cr}.} (\textbf{A}) DFT-calculated band structure of {\Cr}. (\textbf{B}) DFT-calculated band structure of Cs{\V}. In (A) and (B), red arrows mark the energy position of the flat bands (FB) while red circles mark the positions of the VHSs in both {\Cr} and Cs{\V}. (\textbf{C}) Comparison of the DOS of {\Cr} (cyan) and Cs{\V} (orange) whose FB}
\end{figure*}
\addtocounter{figure}{-1}
{
\begin{figure*}  % continued
    \caption{(continued) energy positions are indicated by the red arrows, respectively. (\textbf{D}) Fermi surface of {\Cr} measured with 102eV photons on the left and DFT calculation on the right. Red solid lines mark the 2D projected BZ. Blue arrows denote the light polarization. (\textbf{E}) Same as (D) but at E$_B$ = 0.5eV. The black dashed circles mark the position of the VHS at the M point. (\textbf{F}) The definition and illustration of the orbitals in {\Cr}. (\textbf{G})-(\textbf{J}) band dispersion taken with 114 eV (G)(H) LV  and (I)(J) LH polarization along the $\bar{\Gamma}$-$\bar{K}$-$\bar{M}$-$\bar{K}$ and $\bar{\Gamma}$-$\bar{M}$-$\bar{\Gamma}$ directions. The DFT calculations projected onto the orbitals observable in each measurement geometry according to the selection rules are overlapped on the band dispersions for comparison. Blue arrows denote the polarization direction. The white solid arrow denotes the ${d_{yz}}$ character band position while the white dashed arrow denotes its position in the observation, suggesting a possible orbital-selective band renormalization for ${d_{yz}}$ orbitals. (\textbf{K}) Band dispersions measured with 100 eV photons ({\kz} = 0) along $\bar{\Gamma}$-$\bar{M}$. The measurement geometry and polarization are as marked. (\textbf{L}) EDCs stacking in band dispersions taken with LH polarization at in (K). (\textbf{M}) Same as (L) but taken with LV polarization. Lines of the same colors in (K)-(M) denote high symmetry point positions. 
    }
\end{figure*}}

\begin{figure*}
\includegraphics[width=0.99\textwidth]{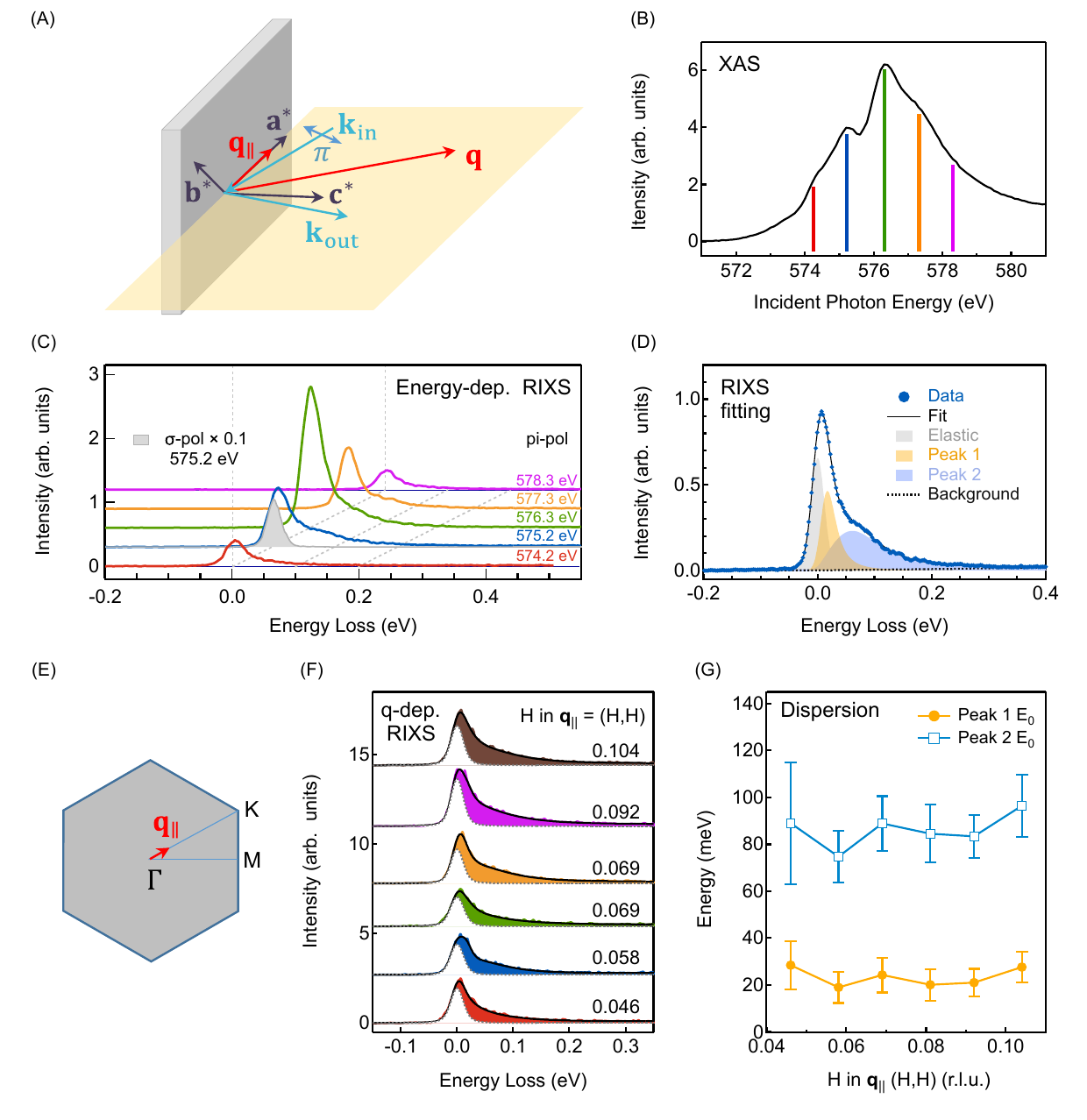}
\caption{\label{fig:fig3} \textbf{RIXS measurements}.
(\textbf{A}) Illustration of RIXS scattering geometry. The scattering plane was perpendicular to the $ab$ plane of \Cr. The incident and scattered wave vectors of X-rays, i.e., $\bf{k}_{\rm in}$ and $\bf{k}_{\rm out}$, are orthogonal. The polarization of incident X-rays was in the scattering plane, i.e., $\pi$-polarized, resulting in RIXS with a cross-polarization geometry. The polarization of scattered X-rays was unresolved. The projection of wavevector change $\bf q$ onto the $ab$ plane is denoted as ${\bf q}_{\|}$. (\textbf{B}) Cr $L_3$-edge X-ray absorption spectrum (XAS) of {\Cr}~recorded at 300~K. Colored vertical bars indicate the X-ray energies used in RIXS measurements. (\textbf{C}) Incident-energy-dependent RIXS with ${\bf q}_{\|}$ along the  ${\Gamma}M$ direction at $25$~K. Spectra in color were recorded with}
\end{figure*}
\addtocounter{figure}{-1}
\begin{figure*}
\caption{(continued) $\pi$-polarized X-rays at selected energies. 
The RIXS spectrum with $\sigma$ polarization, i.e., X-ray polarization perpendicular to the scattering plane, shows the instrumental energy resolution of RIXS. (\textbf{D}) Demonstration of curve fitting for RIXS data analysis. In addition to a linear background, a measured RIXS spectrum was fitted to a spectral profile consisting of three components: one elastic and two electronic excitations. See SM for fitting details.  (\textbf{E}) First Brillouin zone in the $a^{*}b^{*}$ plane of reciprocal space. The red arrow indicates ${\bf q}_{\|}$ of momentum-dependent RIXS measurements. (\textbf{F}) Momentum-dependent RIXS with ${\bf q}_{\|}$ along ${\Gamma}K$ at $25$~K. The energy of incident photons was set to 575.2 eV to optimize the shoulder feature. The black lines plot the elastic components; the colored shades indicate spectral profiles arising from spin excitations. (\textbf{G}) Dispersion of fitted $E_0$ of two spin excitations as a function of in-plane momentum ${\bf q}_{\|}$. }
\end{figure*}

\begin{figure*}
\includegraphics[width=0.99\textwidth]{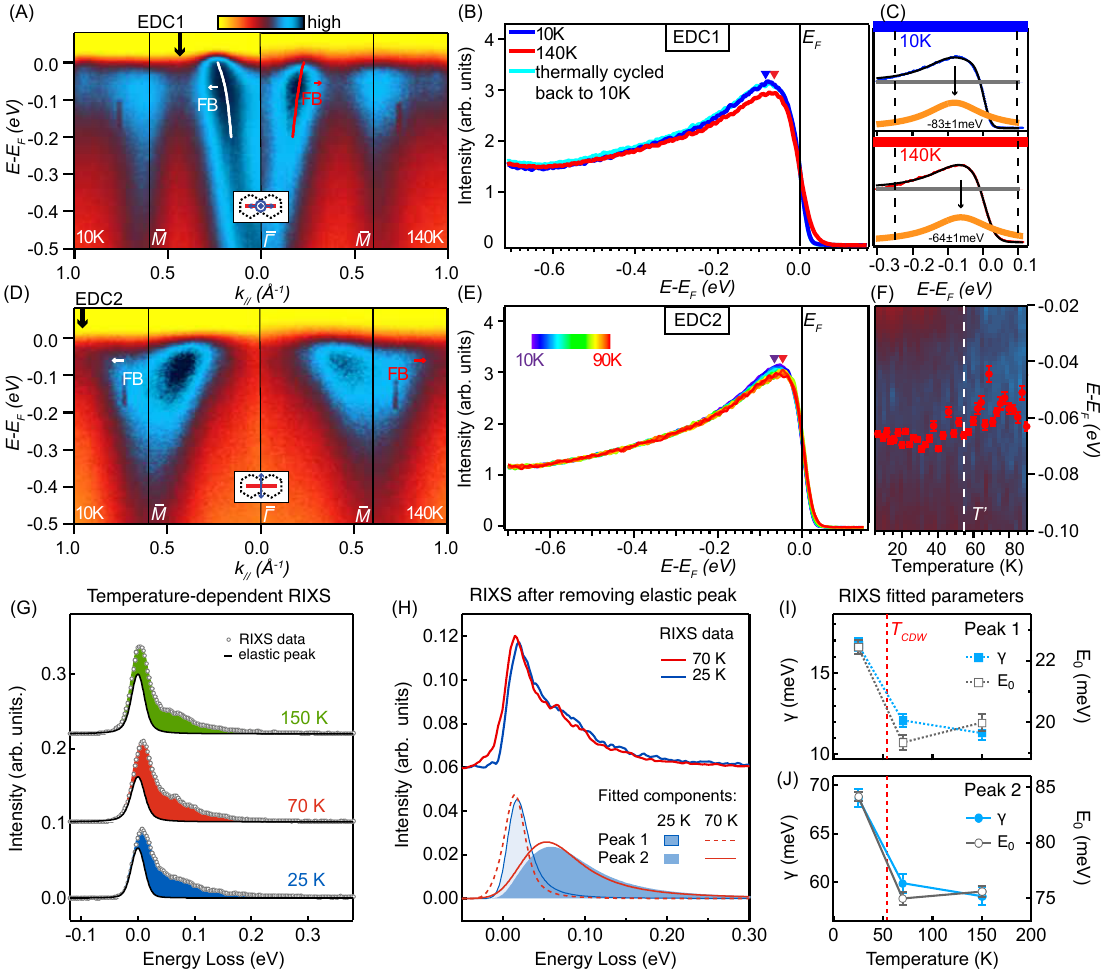}
\caption{\label{fig:fig4} \textbf{Temperature-dependent results of ARPES and RIXS.} (\textbf{A}) Band dispersion taken with 114 eV LH polarization and horizontal slit direction at 10 K (left) and 140 K (right). White (10 K) and red (140K) dots denote the fitted MDC positions (see SM) for the electron pocket at $\Gamma$. The arrows mark a kink in the dispersion indicating hybridization with the FB. (\textbf{B}) EDC1 measured at the denoted position in (A) at 10K, 140K, and then thermally cycled back to 10K. (\textbf{C}) Fitting details for the 10K and 140K EDC1 where a Lorentzian peak (orange) and a constant background (gray) are multiplied by the Fermi-Dirac (FD) function convolved with a fixed Gaussian peak of 40 meV Full-Width-Half-Maximum (FWHM mimicking the experimental resolution). Blue/red dots are the raw data points, the same as (b). Vertical dashed lines mark the fitting range. The dashed arrows denote the fitted Lorentzian peak positions. (\textbf{D}) Same as (A) but measured with LV polarization. (\textbf{E}) EDC2 as marked in (D) measured from 10 K to 90 K. (\textbf{F}) Fitted FB positions}
\end{figure*}
\addtocounter{figure}{-1}
\begin{figure*}
\caption{(continued) from EDCs after dividing the FD function convolved with a Gaussian peak with 40 meV FWHM. The dotted line marks the transition \tc~previously reported~\cite{Liu2023-vk}. (\textbf{G}) Temperature-dependent RIXS. Open circles depict raw RIXS data measured with $\pi$-polarized incident X-rays at an energy of 572.5 eV and temperatures of 25, 70, and 150~K; solid lines plot the fitted elastic components. The shaded areas denote RIXS features arising from spin excitations. (\textbf{H}) RIXS spectra at 25~K (blue) and 70~K (red) after removing the elastic component. The bottom panel compares their fitted components of spin excitations. (\textbf{I-J}) Evolution of fitted parameters $E_0$ and $\gamma$ for temperatures changed across \tc. The vertical dashed line indicates the CDW transition temperature \tc.}
\end{figure*}

\end{document}